\documentclass[10pt,journal,compsoc]{IEEEtran}
  \usepackage{cite}
\ifCLASSINFOpdf
  \usepackage[pdftex]{graphicx}
\else
\fi
\usepackage{amsmath}
\usepackage{algorithmic}

\usepackage{array}
\usepackage{url}

\usepackage{booktabs}
\usepackage{xcolor}

\usepackage{booktabs}
\usepackage[ruled,linesnumbered]{algorithm2e}
\usepackage{hyperref}
\usepackage{amsmath,amssymb,amsfonts}
\usepackage{bm}
\usepackage{multicol}
\usepackage{multirow}
\usepackage{threeparttable}
\usepackage{makecell}

\usepackage{textcomp}
\usepackage[switch]{lineno}

\begin{document}
%
\title{A Data-scalable Transformer for Medical Image Segmentation: Architecture, Model Efficiency, and Benchmark}
%
%
%
%


\author{Yunhe~Gao,
        Mu~Zhou,
        Di~Liu,
        Zhennan~Yan,
        Shaoting~Zhang,
        and Dimitris~N.~Metaxas
\IEEEcompsocitemizethanks{\IEEEcompsocthanksitem Y. Gao, D. Liu and D. Metaxas are with the Computer Science Department, Rutgers University, Piscataway,
NJ, USA.\protect\\
\IEEEcompsocthanksitem M. Zhou and Z. Yan are with SenseBrain Research, Princeton, NJ, USA.\protect
\IEEEcompsocthanksitem S. Zhang is with Shanghai Artificial Intelligence Laboratory, Shanghai, China.}
\thanks{Manuscript received on October 15, 2022. Major revision received on April 4, 2023.}}

\IEEEtitleabstractindextext{%
\begin{abstract}
Transformers have demonstrated remarkable performance in natural language processing and computer vision. However, existing vision Transformers struggle to learn from limited medical data and are unable to generalize on diverse medical image tasks. To tackle these challenges, we present MedFormer, a data-scalable Transformer designed for generalizable 3D medical image segmentation. Our approach incorporates three key elements: a desirable inductive bias, hierarchical modeling with linear-complexity attention, and multi-scale feature fusion that integrates spatial and semantic information globally. MedFormer can learn across tiny- to large-scale data without pre-training. Comprehensive experiments demonstrate MedFormer's potential as a versatile segmentation backbone, outperforming CNNs and vision Transformers on seven public datasets covering multiple modalities (e.g., CT and MRI) and various medical targets (e.g., healthy organs, diseased tissues, and tumors). We provide public access to our models and evaluation pipeline, offering solid baselines and unbiased comparisons to advance a wide range of downstream clinical applications.
\end{abstract}

}

\maketitle

\IEEEdisplaynontitleabstractindextext

%
\IEEEpeerreviewmaketitle

\IEEEraisesectionheading{\section{Introduction}\label{sec:introduction}}

%
%
%
%
\IEEEPARstart{S}{emantic} segmentation is essential in medical image understanding and analysis by parsing raw image data into structured and meaningful categories. These segmented outcomes can benefit the entire clinical workflow, including disease diagnosis \cite{ de2018clinically,devunooru2021deep,shen2015multi,ding2022spatially}, quantitative assessment\cite{van1999quantification,kickingereder2019automated}, treatment planning\cite{nestle2005comparison,nikolov2018deep}, and prognostic monitoring\cite{mitra2015medical}. With the rapid growth of image data, developing data-centric segmentation algorithms is essential to expedite disease detection, reduce inter-reader variability, and enhance diagnostic efficiency in healthcare systems. To achieve this, data-centric segmentation approaches must overcome challenges related to limited data availability, complex anatomy modeling, and algorithm robustness on unseen data. Current research efforts have not yet comprehensively addressed these challenges.

\begin{figure}[]
    \centering
    \includegraphics[width=0.42\textwidth]{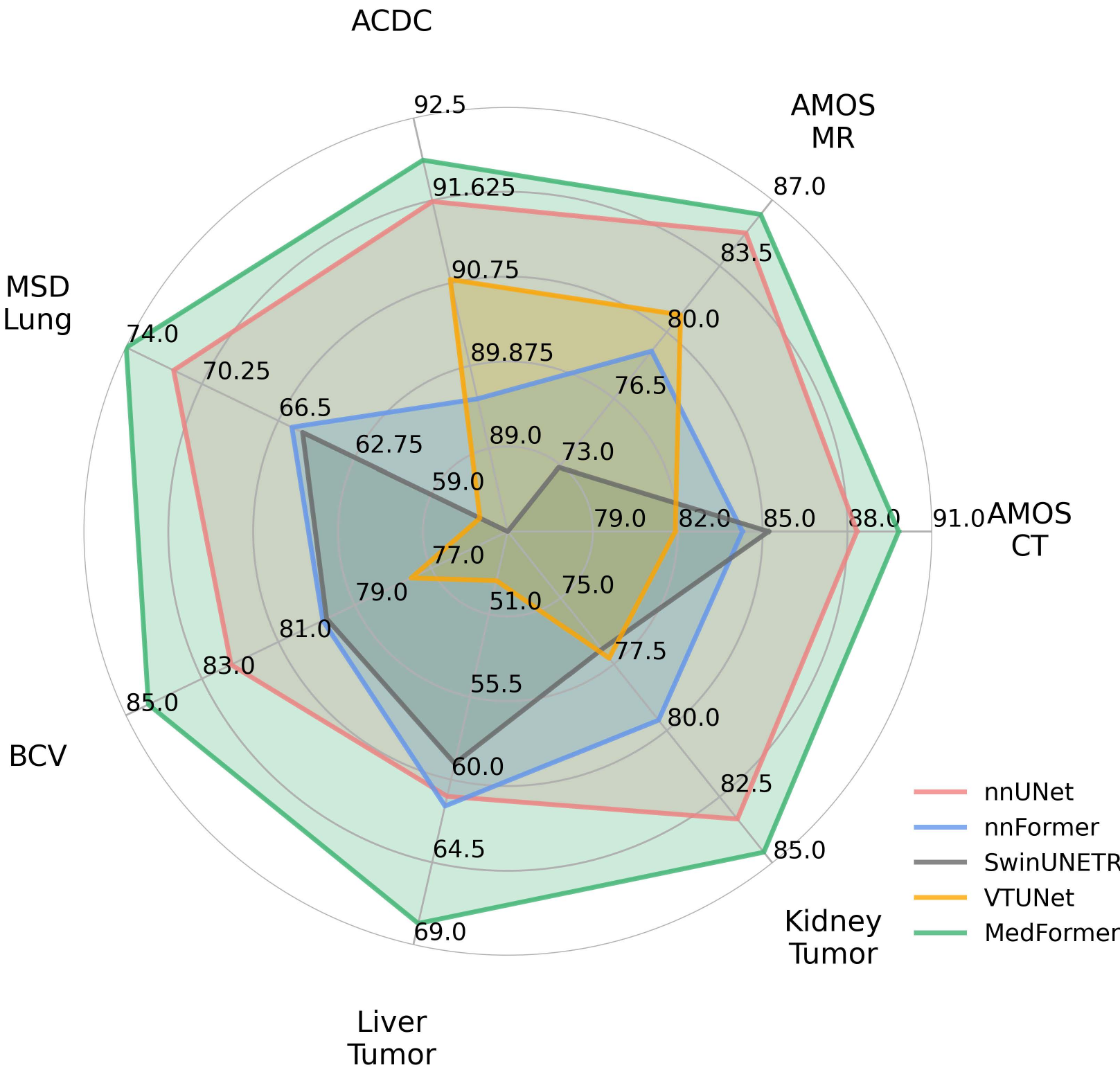}
    \caption{MedFormer exhibits outstanding generalizability across seven diverse public datasets, encompassing various modalities, target structures, and target sizes. While other Transformer-based models display good performance on certain specific datasets, they fail to achieve the same level of generalizability of nnUNet or MedFormer.}
    \label{fig:cover_fig}
\end{figure}

Medical image segmentation has made significant progress due to the advancements in deep neural networks (DNNs) with improved visual representation capabilities. Convolutional neural networks (CNNs) such as U-Net\cite{ronneberger2015u} and its variants\cite{zhou2018unet++,oktay2018attention,gao2019focusnet} evolve to address on a wide range of segmentation tasks\cite{isensee2021nnu}. The recent surge of Transformers, with attention as the key computational primitive, has been proven successful in natural language processing (NLP)\cite{devlin2018bert,radford2019language} and computer vision (CV)\cite{dosovitskiy2020image,liu2021swin}. The rationale for the self-attention mechanism is a double-edged sword in medical tasks. The self-attention mechanism in Transformers offers exceptional global modeling capability on intricate interactions under large-scale training\cite{devlin2018bert,dosovitskiy2020image,radford2021learning}. However, medical tasks face unique challenges \cite{willemink2020preparing}, including data acquisition, annotation cost, and disease diversity. Transformers without inductive bias can struggle to learn from scratch in a low-data regime on disease-specific tasks, and the ImageNet pretrained weights do not transfer well due to the significant domain gap. Moreover, the canonical self-attention has quadratic complexity concerning the input sequence length, making it challenging to balance computation and fine-grained details for high-resolution inputs, especially for 3D images. Although several works have proposed Transformer models for medical image segmentation \cite{chen2021transunet,peiris2021volumetric,hatamizadeh2021unetr,cao2021swin,zhou2021nnformer,hatamizadeh2022swin} and demonstrated satisfactory performance on specialized tasks via tailored designs on architecture or training and testing methodologies, these models often fail to generalize as effectively as UNet \cite{isensee2021nnu}, and in some cases, yield suboptimal performance, as illustrated in Fig. \ref{fig:cover_fig}. This phenomenon has also been corroborated in recent research, such as \cite{ji2022amos, saikat2023mednext}.
Hence, it is crucial to systematically benchmark Transformers and CNNs, and develop a data-scalable Transformer model to accommodate the diverse requirements of medical image analysis tasks.

\begin{figure*}[tb!]
    \centering
    \includegraphics[width=\textwidth]{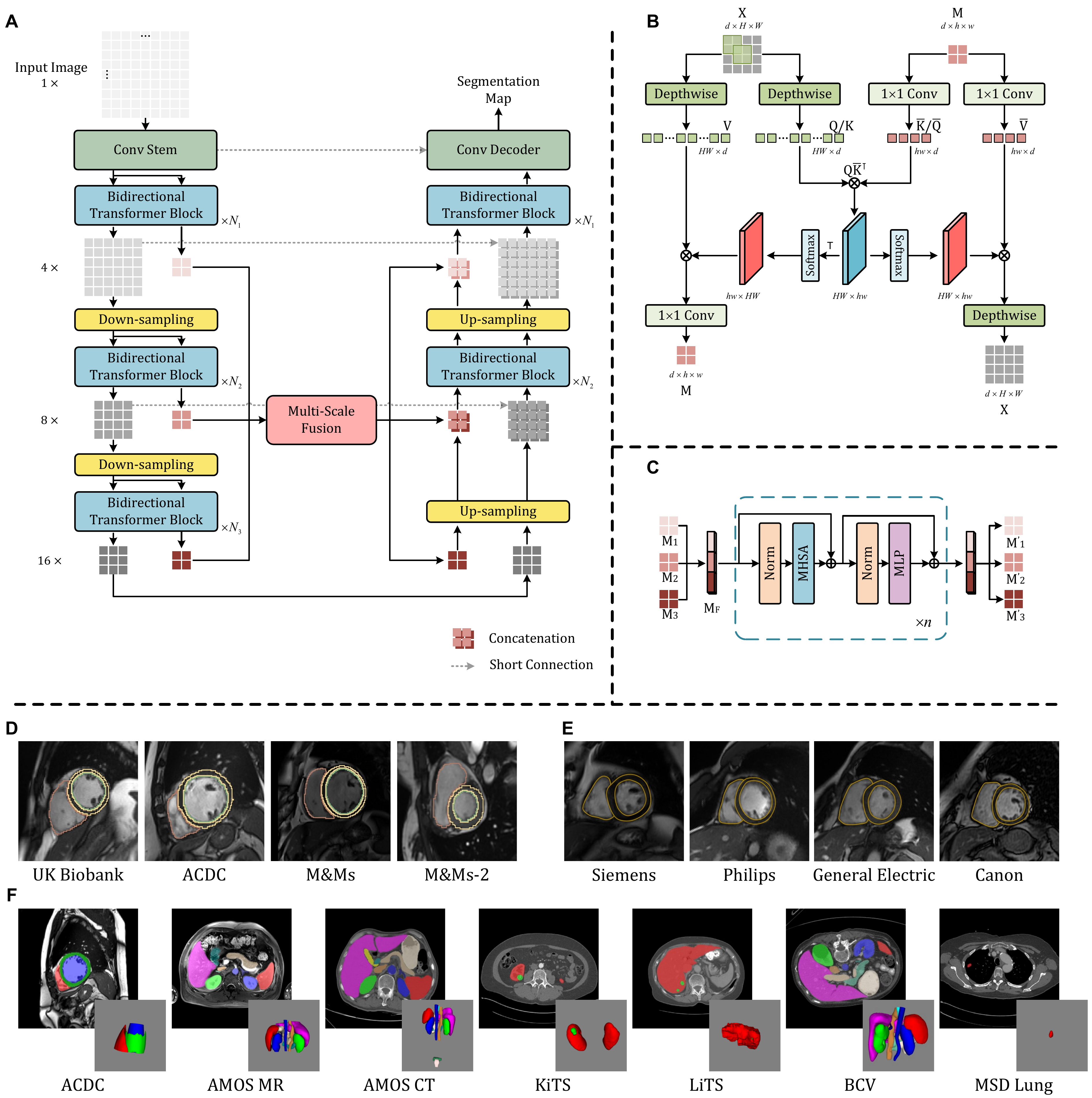}
    \caption{(A): The architecture of MedFormer. (B): The proposed efficient bidirectional multi-head attention (B-MHA) that reduces the complexity of conventional self-attention to be linear. (C): The proposed multi-scale semantic map fusion module. (D): Example images of the collected large cardiac MRI dataset. (E): Example images of the testing data of M\&Ms dataset that comes from four vendors, where domain shift in terms of visual appearance exists among vendors. The testing data samples are prepared for the evaluation of both model performance and robustness. (F): Example  images of seven widely-used datasets that cover diverse tasks, including multiple modalities (e.g., CT and MRI) and various medical targets (e.g., healthy organs, diseased tissues, and tumors).}
    \label{fig:framework}
\end{figure*}

In this study, we develop a hybrid Transformer model, named \textbf{Med}ical Trans\textbf{Former}: \textbf{MedFormer} (Fig. \ref{fig:framework}), for 3D medical image segmentation. Unlike recent endeavors \cite{wu2021cvt,graham2021levit,yuan2021tokens} attempting to reduce training data requirements on the relatively 'small' ImageNet1k dataset\cite{ILSVRC15} (1.28M images), MedFormer can be trained from scratch even on extremely small medical datasets without relying  on pre-training weights. We introduce a desirable inductive bias through the depth-wise separable convolution in the projection and feed-forward network within the Transformer blocks. Moreover, a core contribution of our work is the efficient bidirectional multi-head attention (B-MHA), which eliminates redundant tokens via low-rank projection and reduces the quadratic complexity of conventional self-attention to a linear level. Unlike window-based self-attention\cite{liu2021swin} or decomposing attention\cite{huang2019ccnet,wang2020axial}, the proposed B-MHA can directly model long-range relationships and empowers MedFormer to extract global relations on high-resolution token maps, thereby facilitating fine-grained boundary modeling. Moreover, a semantically and spatially global multi-scale fusion mechanism is incorporated to augment segmentation while incurring negligible computational overhead. Compared with preliminary studies\cite{chen2021transunet,cao2021swin,hatamizadeh2021unetr,peiris2021volumetric,zhou2021nnformer,hatamizadeh2022swin}, our MedFormer exhibits data-scalability, efficiency, and generalizability by demonstrating superior performance on the tiny-scale data without any pre-training weights, while also presenting capacity advantages on the large-scale data regime.

On a collected large cardiac MRI dataset (3,176 3D images) and seven widely-used public datasets with multiple modalities, diverse target structures and size (see in Fig. \ref{fig:framework} (D-F)), we perform extensive experiments across data scale, 2D and 3D settings, emphasizing a comprehensive and unbiased benchmark of state-of-the-art CNNs and Transformer models. Solid baselines are built by training all models within the same framework and training techniques (the codebase is available \footnote{https://github.com/yhygao/CBIM-Medical-Image-Segmentation}). We show that MedFormer achieves superior performance with potential generalizability as a strong segmentation backbone in medical image analysis.

\section{Related Work}

We discuss related topics of medical image segmentation from three major aspects: vision Transformer-based models, efficient attention, and multi-scale feature fusion.

\subsection{Vision Transformer}

Transformer models, with the attention mechanism as the core operator, are emerging in vision tasks and demonstrating promising performance and generalization on image classification via large-scale training \cite{vig2019multiscale,dosovitskiy2020image,tay2021synthesizer}. The self-attention module calculates an all-to-all attention matrix to adaptively derive representations from image tokens, inherently possessing a global receptive field for modeling long-range relationships.  Most improvements of vision Transformer focus on two challenges in the practical application: 1. Substantial demand for training data: Owing to the absence of inductive bias \cite{dai2021coatnet}, ViT relies heavily on large-scale training and remains inferior to CNNs when training data is scarce. 2. Quadratic complexity of self-attention: This complexity presents difficulties in handling lengthy token sequences\cite{liu2021swin}.  These challenges are particularly pronounced in medical image tasks, where training data is often exceedingly limited due to privacy concerns, data acquisition difficulties, and high annotation costs. Additionally, medical images are typically high-resolution 3D images, resulting in exceptionally long token sequences.

Several works attempt to introduce inductive bias into Transformer to reduce the requirement of training data. CvT \cite{wu2021cvt} uses a hierarchical structure and replaces the linear embedding and projection with convolutional embedding and projection, which brings convolution inductive bias to the ViT architecture. CoAtNet \cite{dai2021coatnet} presents a family of hybrid models that unify depthwise convolution and self-attention via relative attention, showing that vertically stacking convolution layers and attention layers in a principled way can improve generalization, capacity and efficiency. These two methods can outperform ViT when training on relatively small ImageNet-1k dataset.

\subsection{Efficient Attention}
Due to the quadratic complexity of self-attention with respect to the input sequence length, ViT has to patchify images into tokens, discarding all structural information within the patch. This aggressive down-sampling design can be troublesome in dense prediction tasks such as segmentation, as detailed information is lost. To reduce the computation complexity, recent works develop three main directions: (1) Local window self-attention. SwinTransformer \cite{liu2021swin} introduces the locality of convolution into self-attention via non-overlap window-based multi-head self-attention (W-MSA).  (2) Decomposing attention. CCNet \cite{huang2019ccnet} proposes a criss-cross module to decompose 2D attention to two 1D attention. This idea is also applied in \cite{wang2020axial}. (3) Low-rank projection. As images are highly-structured data, redundancy exists among tokens. UTNet \cite{gao2021utnet} and concurrent work CvT \cite{wu2021cvt}, PvT \cite{wang2021pyramid} made an effort to reduce the tokens in key and value to improve efficiency. RegionViT \cite{chen2021regionvit} introduced regional tokens to extract global information. Concurrent work DualViT \cite{yao2022dual} proposed a semantic pathway to compress token vectors into global semantics.

Despite significant complexity reduction in the first two directions, locality or decomposition introduces limitations to the receptive field and impairs the ability to directly model long-range relationships. For example, to propagate information across windows, SwinTransformer \cite{liu2021swin} has to apply two shifted-window self-attentions in the consecutive layers. CCNet \cite{huang2019ccnet} requires to recurrently stack multiple criss-cross modules to aggregate full dependencies. In our study, the proposed MedFormer lies in the third direction, where the key effort is placed on retaining useful tokens while eliminating redundant tokens.

There are attempts to apply Transformer in the medical image segmentation field. TransUNet \cite{chen2021transunet} and UNETR \cite{hatamizadeh2021unetr} add 2D or 3D convolutional decoder to ViT-like encoder into medical image segmentation. SwinUNet \cite{cao2021swin} and VT-UNet \cite{peiris2021volumetric} proposed pure Transformer model based on SwinTransformer for 2D or 3D segmentation. nnFormer \cite{zhou2021nnformer}, SwinUNETR \cite{hatamizadeh2022swin} and HiFormer \cite{heidari2023hiformer} use Swin-like hybrid architectures.  These models either need to be initialized with pre-trained weights on large-scale natural image datasets or longer training epochs, otherwise they likely achieve inferior performance on medical image datasets.

\subsection{Multi-scale feature fusion}

Multi-scale feature fusion has been proven as an effective strategy for dense prediction tasks \cite{cai2016unified,chen2018cascaded,chen2017deeplab}. One approach is resampling input images into a multi-resolution input pyramid, processing them through multiple networks, and aggregating the output \cite{tompson2015efficient,lin2021ds,chen2021crossvit}. Alternatively, feature pyramids, such as UNet \cite{ronneberger2015u} and FPN \cite{lin2017feature}, gradually fuse features by concatenating up-sampled high-level features from the decoder with low-level features via shortcut connections from the encoder. Researchers have also explored adding more connections between high- and low-level representations, as exemplified by UNet++ \cite{zhou2018unet++} and HRNet \cite{wang2020deep}. Atrous spatial pyramid pooling \cite{chen2017rethinking} can also achieve multi-scale fusion by capturing multi-scale features within the same level.

These approaches are typically considered as local fusion since they primarily use convolution to fuse semantic features of different scales locally, without accounting for global information. For instance, the features of an object in one image corner will not contribute to understanding another similar object in the opposite corner due to limited receptive fields. Recently, several methods have been proposed for global fusion, such as CoTr \cite{xie2021cotr}, which fuses flattened multi-scale feature maps from the CNN encoder using attention. However, CoTr's fusion process is computationally demanding due to the long sequence length. In contrast, our proposed method focuses on a global multi-scale fusion of image semantic features through the introduction of a semantic map, enhancing fine-grained segmentation with minimal computational overhead.

\section{Method}

In this section, we elaborate on the core ideas of MedFormer by addressing three major questions. First, how do we design a unified model that has a large model capacity while does not require an extensive amount of training data? Second, how can we reduce the complexity of attention for high-resolution inputs without degrading performance? Finally, what is a better design for fusing multi-scale information in semantic segmentation? The proposed techniques are applicable for both 2D and 3D settings, while we use 2D formulas for simplicity in the following sections.

\subsection{Preliminary}

The canonical Transformer is built upon the multi-head self-attention (MHSA) module and feed-forward network (FFN) \cite{vaswani2017attention}. For MHSA in vision tasks, given a representation map $X\in \mathcal{R}^{d\times H\times W}$, where $H$,$W$ are the spatial height, width and $d$ is the number of channels. Every pixel is treated as a token. The token map is flattened to a sequence as the input of the Transformer block: $\textbf{X}\in \mathcal{R}^{n\times d}$ (bold for flattened 1D sequence, while regular for 2D token map), where $n=HW$ is the sequence length. Three linear transformations are used to project $\textbf{X}$ to query, key and value embeddings: $\textbf{Q, K, V}\in \mathcal{R}^{n\times d}$.  The scaled dot-product attention used by Transformer is given by:

\begin{equation}
    {\rm Attention}(\textbf{Q, K, V})=\underbrace{{\rm softmax}(\frac{\textbf{QK}^{\mathsf{T}}}{\sqrt{d}})}_{A}\textbf{V}
\end{equation}

$A\in \mathcal{R}^{n\times n}$ is often called the attention matrix that measures the similarity of each token-pairs as weights for aggregate context information from value embedding. Transformer uses multi-head self-attention that projects the query, key, and value embeddings to multiple representation sub-spaces for attention computation and then concatenates the outputs of multiple heads together as the final output. Without loss of generality, we omit the multi-head in all formulas for simplicity. The feed-forward network is a position-wise two-layer perceptron that consists of two linear layers and an activation function that operates separately and identically on each position. Overall, the two-layer perceptron works as a feature transformation layer that increases model capacity by introducing non-linear transformation.

\subsection{Introducing Convolutional Inductive Bias}

The position-wise linear projection transforms the tokens element-wise without interacting with each other, making the Transformer to be permutation invariant. All local structure information that is vital for images are not taken into account in this step. Although adding positional encoding allows the Transformer to learn position relationships, learning from scratch demands a tremendous amount of training data, which is a crucial pain point of medical image analysis. Therefore, we propose to introduce the desirable inductive bias of convolution to the projection of the attention and the feed-forward network, see in Fig. \ref{fig:framework} (B). Given a 2D token map $X$, a convolution with kernel size $k$ is implemented to project the feature map into different spaces and is then flattened into 1D for subsequent attention computation as query, key, or value, formulating as:

\begin{equation}
    \textbf{Q}/\textbf{K}/\textbf{V}= {\rm Flatten}({\rm Conv}(X, k))
\end{equation}

We use the depth-wise separable convolution\cite{chollet2017xception} as an efficient version of convolution implemented by: $\rm depthwise\ conv \rightarrow pointwise\ conv$, where the depth-wise convolution gathers the spatial information while the point-wise convolution gathers along the channel dimension. For FFN, we adopt a similar modification. Given token sequence $\textbf{X}$ after the attention module, we first reshape them back to 2D and transform them with convolutional blocks:

\begin{equation}
    X = {\rm ConvBlock}({\rm Reshape2D}(\textbf{X}), k)
\end{equation}
 We use the MBConv\cite{sandler2018mobilenetv2} as the convolutional blocks, which consists of: $\rm depthwise\ conv \rightarrow activation \rightarrow pointwise\ conv$. The proposed convolutional projection and feed-forward network are a generalized version of the origin Transformer design, which can be implemented using $1\times 1$ convolution layer in both modules.

\subsection{Efficient Attention} The computation bottleneck of the vision Transformer comes from the attention module. The dot-product of two $n\times d$ matrices leads to $O(n^2d)$ complexity. Typically, the sequence length $n$ is much larger than $d$ when the resolution of the token map is high, especially for 3D tasks with a large volume size, thus dominating the self-attention computation and making it infeasible to apply self-attention in high-resolution token maps.  
As images are highly structured data, most pixels in high-resolution feature maps within local footprint share similar semantic meanings, the all-to-all attention is highly inefficient and redundant. From a theoretical perspective, self-attention is essentially low rank for long sequences\cite{wang2020linformer}, which indicates that most information is concentrated in the largest singular values. Inspired by this finding, UTNet\cite{gao2021utnet} proposed an efficient self-attention mechanism by reducing the number of tokens in key and value through sub-sampling. Similar ideas are also applied in concurrent work\cite{wu2021cvt,wang2021pyramid}. The main idea is using two low-rank projections to the key and value: $\mathbf{K, V}\in \mathcal{R}^{n\times d}$ to: $\overline{\mathbf{K}}, \overline{\mathbf{V}}\in \mathcal{R}^{l\times d}$, where $l=hw\ll n$, $h$ and $w$ are the reduced size of the token map after low-rank projection. The efficient self-attention is:

\begin{equation}
    {\rm Attention}(\textbf{Q}, \overline{\textbf{K}}, \overline{\textbf{V}})=\underbrace{{\rm softmax}(\frac{\textbf{Q}\overline{\textbf{K}}^{\mathsf{T}}}{\sqrt{d}})}_{\overline{A}:n\times l}\underbrace{\overline{\textbf{V}}}_{l\times d}
\end{equation}

By doing so, the computational complexity is reduced to $O(nld)$, which is linear to the input sequence length $n$.

\subsection{Bidirectional Attention (B-MHA).} 
\begin{figure}[t]
    \centering
    \includegraphics[width=0.4\textwidth]{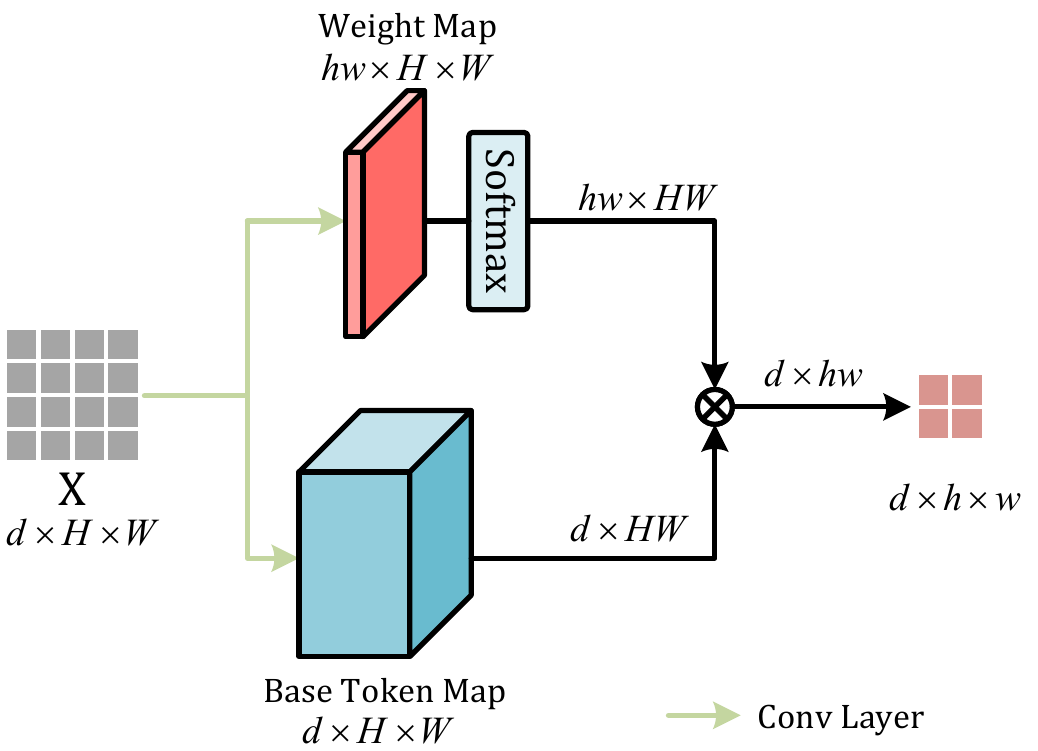}
    \caption{Illustration of the initial semantic map generation.}
    \label{fig:init_map}
\end{figure}

Although the mentioned efficient attention substantially reduces computation, it remains sub-optimal due to their usage of simple linear transformations, such as interpolation or stride convolution, for low-rank projections. These linear and local operations lack a holistic perspective to preserve the most informative tokens when the compression ratio is high.  Therefore, we proposed a bidirectional multi-head attention module (B-MHA) that learns to effectively project the full image token map into a concise semantic map that stores the holistic semantics by a non-linear dimensionality reduction. To be specific, an initial semantic map with a small spatial size is projected at each level and subsequently refined by the B-MHA module. The initial semantic map generation process is shown in Fig. \ref{fig:init_map}). Given a token map $X$ with size $d\times H\times W$, where $d$ is the channel number, $H$ and $W$ are the spatial height and width, it is projected with two convolutional layers to a weight map and a base token map. The weight map has a channel number of $hw$, where $h\ll H$ and $w\ll W$ are the predefined sizes of the semantic map. The weight map is then flattened for softmax computation, and works as the weight to aggregate semantic information from the base token map by computing the matrix product. 

The B-MHA module has two inputs, see in Fig. \ref{fig:framework} (B), one is the full image token map $X$ from the previous layer, while another is the semantic map $M$. The $X$ and $M$ are projected to $\textbf{Q}/\textbf{K}/\textbf{V}$ and $\overline{\textbf{Q}}/\overline{\textbf{K}}/\overline{\textbf{V}}$ respectively for a cross-attention. The $X$ is projected with depthwise separable convolution while $M$ is projected with $1\times 1$ convolution. As $M$ has a much smaller size and each element in $M$ has valuable semantic information, the padding in $3\times 3$ depthwise convolution will introduce noise to the semantic map. To reduce the computation and memory consumption, query and key of $X$ and $M$ are shared. As the the dot product of query and key measures the similarity of a token-pair, which is symmetrical, we can reuse it for compute the attention matrix to aggregate context for both $X$ and $M$ by simply transposing the dot product matrix (before softmax):

\begin{equation}
    \begin{split}
        X'=Attention(\textbf{Q}, \overline{\textbf{K}}, \overline{\textbf{V}}) &= {\rm softmax}(\frac{\textbf{Q}\overline{\textbf{K}}^{\mathsf{T}}}{\sqrt{d}}) \overline{\textbf{V}}   \\
        M'=Attention(\overline{\textbf{Q}}, \textbf{K}, \textbf{V}) &= {\rm softmax}(\frac{\overline{\textbf{Q}} \textbf{K}^{\mathsf{T}}}{\sqrt{d}})\textbf{V}  \\
        \overline{\textbf{Q}} \textbf{K}^{\mathsf{T}} &= (\textbf{Q}\overline{\textbf{K}}^{\mathsf{T}})^{\mathsf{T}}
    \end{split}
\end{equation}

B-MHA's non-linear update of the semantic map enables more effective dimensionality reduction compared to linear low-rank projection. The semantic map preserves and continuously refines low-dimensional information, serving as a holistic summary of the high-dimensional token map. This allows the attention module to capture context information with significantly reduced computation. The depthwise separable convolution in the projection and FFN excels at capturing local responses, making the B-MHA Transformer block adept at modeling both local and global relationships.

B-MHA's low-rank projection not only enhances computational efficiency but also reduces optimization difficulty and introduces additional regularization implicitly. The limited number of meaningful semantics in medical images, such as different organs and tissues, necessitates the low-rank projection to avoid overfitting on task-irrelevant features and ensure the model learns efficient and robust data representations. This results in improved generalization on unseen data, particularly for small datasets with limited training data.

\subsection{Global Multi-scale Semantic Map Fusion}

Multi-scale fusion plays a vital role in dense prediction tasks to combine the high-level semantic and low-level detailed information. The semantic map in B-MHA is naturally suitable for multi-scale fusion with a minimal computation overhead. See in Fig. \ref{fig:framework} (C), given 2D semantic maps from multiple scales: $M_1, M_2, \dots, M_n$, we first flatten them and concatenate them together into a long 1D token sequence $\textbf{M}_F$. 
The sequence $\textbf{M}_F$ contains all tokens from all scales and is then fed into conventional Transformer blocks for multi-scale semantic fusion. The fused sequence is then chunked and reshaped back to 2D semantic maps. Unlike previous approaches fuse multi-scale features locally, such as fusing with resized multi-scale feature\cite{zhou2018unet++} or with atrous spatial pyramid pooling\cite{gao2021focusnetv2}, the proposed approach propagates information across all tokens at every scale via the all-to-all attention to form a semantically and spatially global multi-scale fusion.

\subsection{MedFormer}

After introducing the key components, we stack them together to become a powerful segmentation model, see in Fig. \ref{fig:framework} (A). MedFormer utilizes a convolutional stem, consisting of several convolutional residual blocks\cite{he2016identity} and down-sampling layers to embed input images to $4\times$ down-sampling token maps. Such embedding eliminates the structural information loss and accepts arbitrary input size. A hierarchical representation is built with the proposed B-MHA Transformer blocks and down-sampling layers implemented with patch merging layer\cite{liu2021swin}. Specifically, along with the full-size token map, a holistic semantic map is generated\cite{liu2021holistically} at every scale as the input of B-MHA (details can be provided in the supplementary). The hierarchical multi-scale information within the semantic map is fused through the proposed fusion module. MedFormer gradually restores the resolution through a series of up-sampling and B-MHA blocks in the decoder with the token maps from the encoder and the corresponding fused semantic map. At last, a convolutional decoder combines the high-resolution feature maps from the convolutional stem to output the final segmentation map. In order to improve the training efficiency of the Transformer component of MedFormer instead of being bypassed by the convolutional stem, we add deep supervision by computing an auxiliary loss on the output of the last B-MHA block in the decoder during training.

\section{Experiments}

We conduct systematic experiments to evaluate various model architectures across diverse settings, categorized into three groups based on objectives. First, using a large cardiac MRI dataset, we examine the influence of data quantity on different models and assess their robustness across multiple vendors. Second, we investigate the models' generalization capabilities using seven public datasets, comparing state-of-the-art CNN and Transformer-based models under a consistent framework. Third, we perform an in-depth analysis of MedFormer's effectiveness. Detailed information about the datasets, compared models, and experiment settings can be found in Appendices A, B, and C, respectively.

\subsection{The large cardiac MRI dataset.} 
Fig. \ref{fig:framework} (D) presents image examples from our collected large cine MRI dataset, including ACDC (100 cases)\cite{bernard2018deep}, M\&Ms (320 cases)\cite{campello2021multi}, M\&Ms-2 (160 cases)\cite{campello2021multi}, and UK Biobank (UKBB) (1,008 cases)\cite{petersen2015uk}. All datasets share identical target annotations for left ventricle (LV), right ventricle (RV), and left ventricular myocardium (MYO). In total, 1,588 cine MRI scans are labeled in end-diastolic (ED) and end-systolic (ES) phases, culminating in 3,176 3D MR images. Notably, the M\&Ms dataset comprises images from four scanner vendors (A: Siemens, B: Philips, C: General Electric, D: Canon), resulting in a visual appearance gap as depicted in Fig. \ref{fig:framework} (E). Consequently, the M\&Ms dataset facilitates both model performance and robustness assessments. We employ the M\&Ms test set for evaluation, containing 170 cases (A: 20, B: 50, C: 50, D: 50), while the M\&Ms training set, combined with the other three datasets, constitutes our training set. It should be noted that the large training set includes a limited number of images from vendor C and none from vendor D. We designate vendor C as a rarely-seen domain and vendor D as a completely unseen domain, both serving for out-of-distribution evaluations.

\begin{table*}[]
    \centering
    \caption{DSC results on the large cardiac MRI dataset with different training data ratios}
    \setlength{\tabcolsep}{6mm}{
    \begin{threeparttable}
    \begin{tabular}{c|c|c|c|c|c|c}
    \toprule
                Arch.   & Models &5\% & 10\% & 40\% & 70\% & 100\% \\ \hline \midrule
    \multirow{6}*{CNN}  & UNet\cite{ronneberger2015u}  & 86.51 & 87.17 & 87.96 & 88.37 & 88.59  \\ \cline{2-7}
                        & Attn UNet\cite{schlemper2019attention} & 86.74 & 87.46 & 88.20 & 88.38 & 88.63 \\ \cline{2-7}
                        & UNet++\cite{zhou2018unet++} & 86.54 & 87.33 & 88.21 & 88.52 & 88.59 \\ \cline{2-7}
                        & ResUNet  & 86.65 & 87.48 & 88.31 & 88.54 & 88.49 \\ \cline{2-7}
                        & R50-UNet\cite{he2016deep}  & 86.42 & 87.48 & 88.39& 88.59 & 88.72 \\ \cline{1-7}
    \multirow{6}*{TFM}  & TransUNet\cite{chen2021transunet}  & 86.53 & 87.56 & 88.33  &88.55 & 88.56\\ \cline{2-7}
                        & TransUNet$\dagger$  & 86.30 & 87.22 & 88.01  & 88.57 & 88.55 \\ \cline{2-7}
                        & SwinUNet\cite{cao2021swin}  & 75.25 & 82.19  & 85.71   & 86.20 & 86.83  \\ \cline{2-7}
                        & SwinUNet$\dagger$ & 86.71 & 86.91 & 87.61 & 87.70 & 88.01  \\ \cline{2-7}
                        & UTNet\cite{gao2021utnet}  & 86.70 & 87.50 & 88.41 &  88.58  & 88.69\\ \cline{2-7}
                        & MedFormer & \textbf{87.72}  & \textbf{87.99}  & \textbf{88.80} &  \textbf{88.92}  & \textbf{89.05} \\

    \bottomrule
   
    \end{tabular}
    
    \begin{tablenotes}
    \footnotesize
    \item $\dagger$ indicates the model is initialized with pre-trained weights on ImageNet.
    \end{tablenotes}
    
    \end{threeparttable}
    \label{tab:cardiac_large}}
\end{table*}

\begin{figure*}[]
    \centering
    \includegraphics[width=0.9\textwidth]{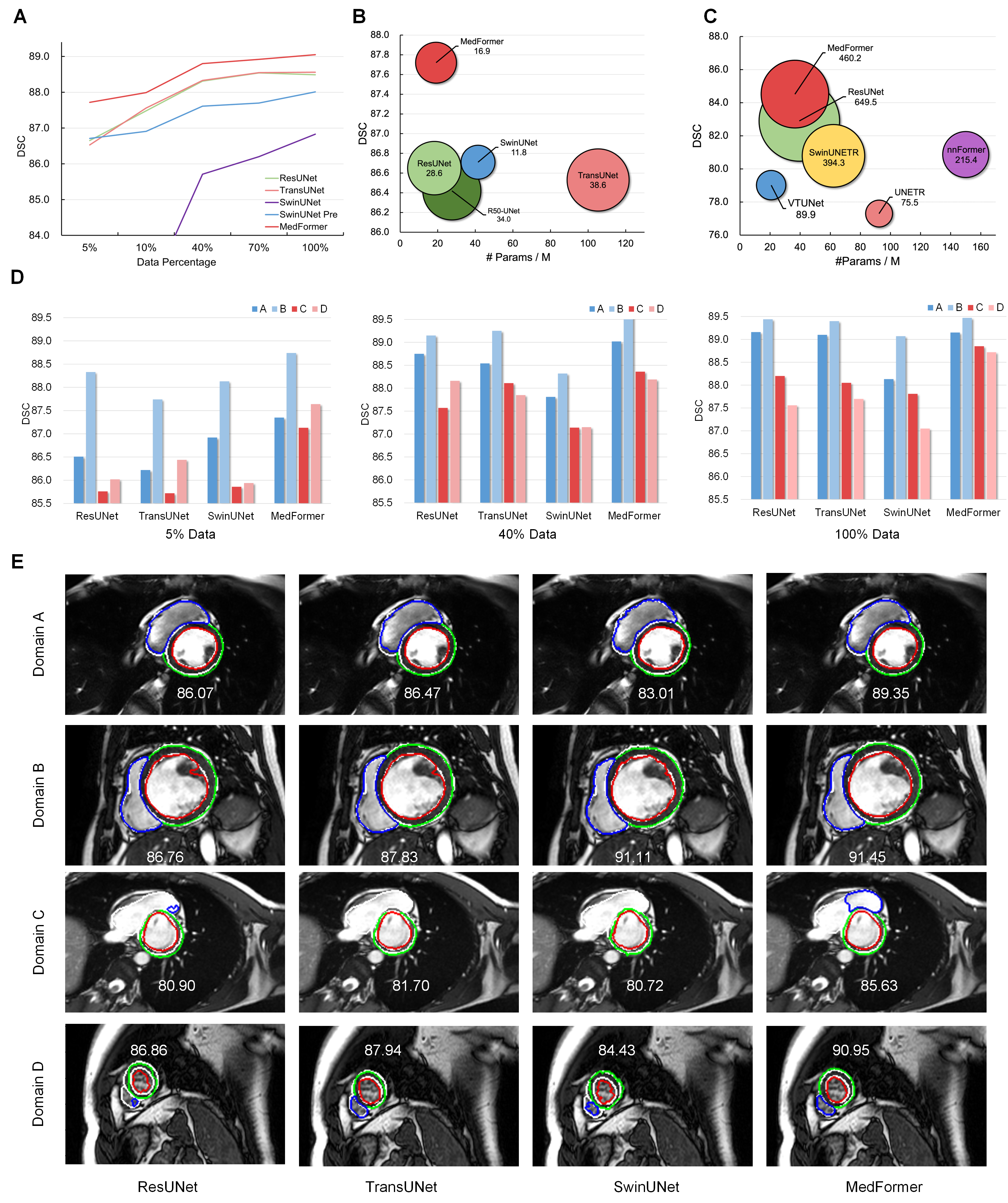}
    \caption{(A): The curve of Dice score coefficient (DSC) v.s. training data percentage in the collect cardiac dataset. (B) and (C): DSC v.s. number of parameters v.s. Flops. y-axis: DSC, x-axis: number of parameters/M, bubble size/number under model: the Flops/G. (B) shows 2D models comparison under 5\% percent data, measured with $256\times 256$ input size. (C) shows 3D models comparison on BCV dataset, measured with $64\times128\times128$ input size. (D): The robustness analysis of models across data scales on four testing vendor domains. A detailed setting description can be found in the method section. (E): Segmentation boundary of models trained with 5\% data on four domains. White: ground truth, Red: LV. Green: MYO. Blue: RV. The numbers on the lower right of the images are the Dice scores.}
    \label{fig:fig2}
\end{figure*}

\subsection{MedFormer is data-scalable.}
MedFormer demonstrates superior performance over other CNNs and vision Transformer models  from limited data to large data. Table \ref{tab:cardiac_large} and Fig. \ref{fig:fig2} (A) show the performance of each model with varying training data ratios (e.g., 5\%, 10\%, 40\%, 70\% and 100\%) of the entire dataset. All model architectures exhibit a data-driven feature, with increased training data-sample inputs consistently leading to higher performance. The improvement from 5\% to 40\% is particularly noticeable. To assess model validity on a small-scale data, we focus on the ratios under 5\% (70 cases) and 10\% (141 cases). In these scenarios, MedFormer exhibits the highest Dice score compared with all strong baselines. For instance, MedFormer under 5\% data even outperforms all other competing models under 10\% training data. Moreover, in large-scale data settings like 70\% (992 cases) and 100\% (1,418 cases), MedFormer exhibits outstanding scalability with respect to data quantity.

Pure Transformer models without pre-training weights exhibit suboptimal performance on small-scale datasets, and ImageNet pre-trained weights provide limited benefits for medical image segmentation due to the substantial domain gap. For instance, SwinUNet without pre-training weights performs considerably worse than other models when using only 5\% and 10\% of the training data. This is attributed to the pure Transformer architecture lacking inductive bias, which leads to a significant data requirement. When utilizing transferred ImageNet22K pre-training weights, the pure Transformer model's performance improves, accompanied by accelerated convergence. For example, the pre-trained SwinUNet marginally outperforms ResUNet in a small dataset setting (5\%). However, owing to the considerable domain gap between natural and medical images, the performance gains from ImageNet22K pre-training weights diminish as the amount of medical training data increases. As illustrated in Fig. \ref{fig:fig2} (A), SwinUNet with pre-trained weights falls behind ResUNet at data scales ranging from 10\% to 100\%. In contrast, MedFormer maintains exceptional results across all scales (5\%-100\%) without requiring any pre-training.

Employing Transformer blocks solely on low-resolution features offers limited advantages for segmentation tasks. The comparison between TransUNet and R50-UNet indicates the usefulness of ViT\cite{dosovitskiy2020image} on the $16\times$ down-sampled feature maps, as seen in Table \ref{tab:cardiac_large}. It is observed that TransUNet achieves performance comparable to R50-UNet across various data ratios. In fact, the coarse modeling of such low-resolution features provides minimal assistance in capturing fine-grained details in segmentation tasks, with the primary contribution to TransUNet's performance likely attributed to the ResNet50 backbone.

MedFormer exhibits enhanced capacity when handling large-scale data. As depicted in Fig. \ref{fig:fig2}(A) for data ratios ranging from 40\% to 100\%, the performance of both ResUNet and TransUNet gradually plateaus, while MedFormer continues to improve. These observations corroborate previous findings indicating the advantages of Transformer architectures when leveraging large-scale data support\cite{dosovitskiy2020image,liu2021swin,dai2021coatnet}. Notably, MedFormer showcases robust data scalability. In the absence of pre-training weights, MedFormer rapidly converges to achieve high performance with small-scale data, and also displays considerable capacity in large-scale settings.

\subsection{MedFormer efficiently handles high-resolution medical data.}

Fig. \ref{fig:fig2} (B) and (C) display the comparison of the number of parameters, performance, and FLOPs. In both 2D and 3D settings, MedFormer achieves leading performance with fewer parameters and moderate computational requirements. This advantage stems from the design of the bidirectional multi-head attention (B-MHA) mechanism. B-MHA promotes linear complexity with respect to input sequence length by minimizing redundancy within token maps, allowing MedFormer to efficiently model long-range relationships in high-resolution token maps for precise boundary delineation. In comparison, the large number of parameters introduced by ViT does not enable TransUNet to significantly outperform R50-UNet. SwinUNet, VT-UNet, nnFormer and SwinUNETR, as SwinTransformer variants, they fail to markedly outperform the CNN baseline due to limited training data and modeling capabilities.

\subsection{MedFormer demonstrates robustness against domain shifts.}

Domain shifts are commonly observed in clinical settings due to inherent differences among scanner vendors, scanning protocols, and image qualities. Assessing model robustness against domain shifts is crucial for determining model validity in real-world applications. Fig. \ref{fig:fig2} (D) displays the performance of models on four domains under different training data ratios, with domains A and B being training domains. MedFormer exhibits exceptional robustness against domain shifts across data scales. With only 5\% of the data volume, MedFormer performs competitively in domains A and B. More importantly, MedFormer maintains a significant advantage (above 87\%) over other baselines on unseen domains C and D, serving as robustness indicators. As the amount of training data increases, the performance of each model in each domain gradually improves; however, the robustness varies significantly. With 100\% of the training data volume, ResUNet and TransUNet perform similarly to MedFormer on domains A and B, but their performance declines notably on domains C and D. MedFormer's robustness advantage can be attributed to B-MHA, which retains the most valuable and salient semantic information in the semantic token map while eliminating irrelevant tokens. Fig. \ref{fig:fig2} (E) visualizes the segmentation outcomes of models on all four domains, further corroborating our findings. Although every model demonstrates decent segmentation on training domains A and B, MedFormer exhibits notable improvements on the challenging unseen domains C and D.

\begin{figure*}[]
    \centering
    \includegraphics[width=0.75\textwidth]{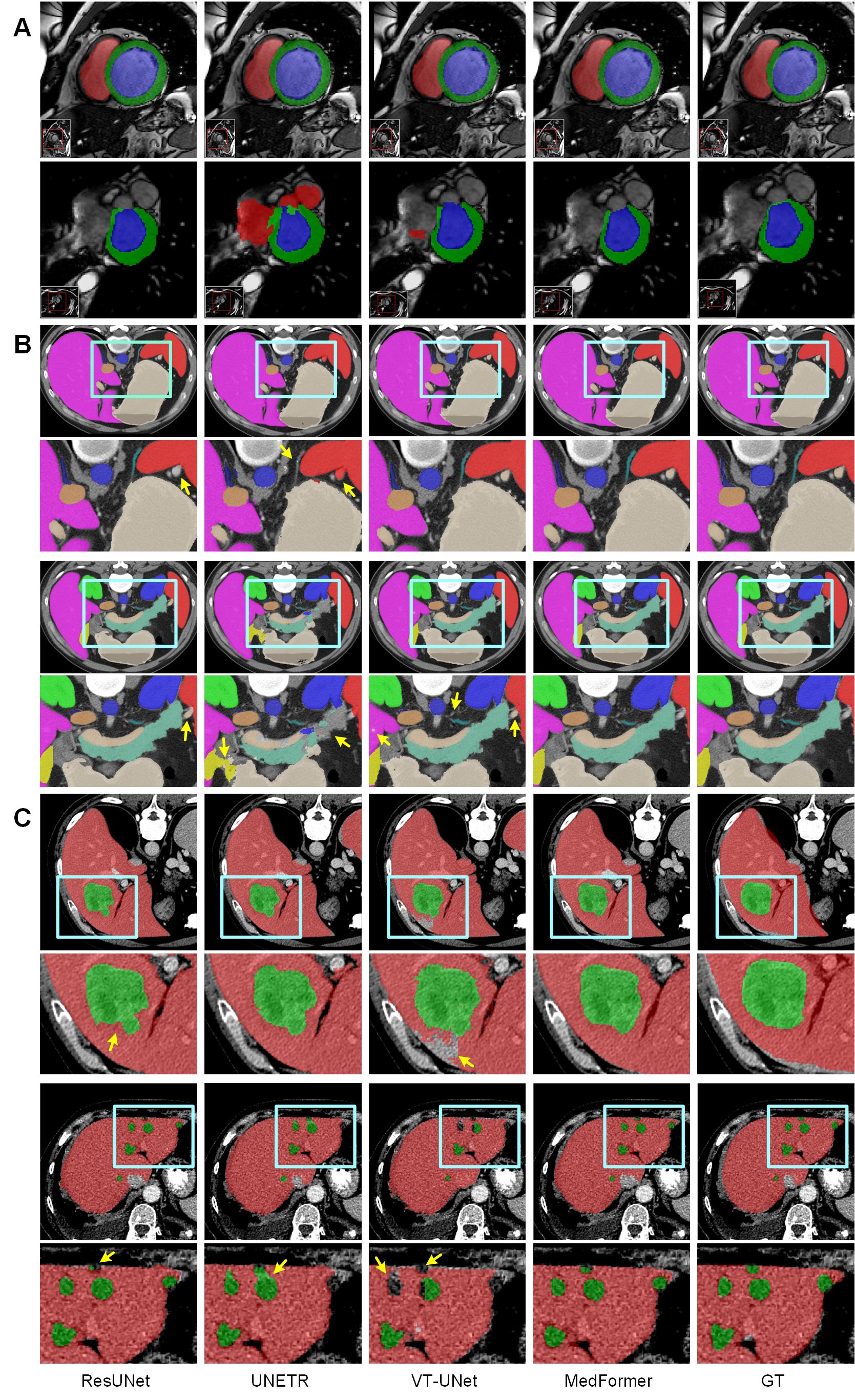}
    \caption{Segmentation visualization. (A): ACDC dataset. (B): BCV dataset, (C): LiTS dataset. Yellow arrows indicate notable and challenging areas of segmentation errors from different models.}
    \label{fig:fig3}
\end{figure*}

\begin{table*}[h]
\renewcommand\arraystretch{1.2}
    \centering
    \caption{Generalization comparison on seven public datasets.}
    \setlength{\tabcolsep}{4mm}{
        \begin{threeparttable}    
    \begin{tabular}{c|c|c|c|c|c|c|c|c}
    \toprule
                & ACDC  & BCV    & LiTS  & KiTS   & AMOS     & AMOS  & MSD   & Avg\\ 
                &       &        & Tumor & Tumor  & CT       & MR    & Lung  & \\\hline
    nnUNet      & \underline{91.79} & 82.79  & 62.53 & \underline{83.67}  & 88.87    & 85.59 & \underline{72.11} & 81.05 \\\hline
    ResUNet     & 91.30 & 82.94  & 63.42 & 83.28  & 88.92    & 85.79 & 71.26 & 80.98 \\\hline
    Attn UNet   & 91.44 & 83.13  & 64.07 & 83.63  & \underline{89.09}    & 85.84 & 71.37 & 81.23 \\\hline
    DeepLabv3+  & 91.51	& \underline{83.27}	 & \underline{64.19} & 83.61  & 89.02	 & \underline{85.96} & 71.43 & \underline{81.28}\\\Xhline{2\arrayrulewidth}
    UNETR       & 87.51 & 77.30  & 53.02 & 70.52  & 80.33    & 77.30  & 55.21 & 71.02 \\ \hline
    SwinUNETR   & -     & 80.79  & 61.10 & 78.55  & 86.37    & 75.70 & 67.05 & -\\\hline
    VT-UNet$\dagger$& 91.13	& 79.02	 & 53.14 & 78.82  & 83.73	 & 82.13 & 60.09 & 75.43\\\hline
    nnFormer    & 90.12 & 80.87  & 62.95 & 80.69  & 85.63    & 80.60 & 67.47 & 78.33 \\\hline
    MedFormer   & \bf92.14	& \bf84.52	 & \bf68.06 & \bf84.47  & \bf90.11	 & \bf86.37 & \bf73.97 & \bf82.83\\

    \bottomrule
    \end{tabular}
    \begin{tablenotes}
    \footnotesize
    \item $\dagger$ indicates the model is initialized with pre-training weights on ImageNet22K. All
     models are 3D models.
    \end{tablenotes}

    \end{threeparttable}
    \label{tab:pub}}
\end{table*}

\subsection{MedFormer exhibits superior generalizability across diverse medical tasks.}

\begin{table}[h]
\footnotesize
    \centering
    \caption{Information of the seven datasets}
    \setlength{\tabcolsep}{1.5mm}{
        \begin{threeparttable}    
    \begin{tabular}{c|c|c|c|c}
    \toprule
    Name & Modality & \#Cls & \#Data & Note \\\hline
    ACDC \cite{bernard2018deep} & cineMRI  &  4   & 100    & Cardiac  \\\hline
    LiTS \cite{bilic2019liver} & CT       & 3     & 131   & Liver tumor \\\hline
    KiTS \cite{heller2019kits19} & CT       & 3   & 210    & Kidney tumor \\\hline
    MSD Lung \cite{antonelli2022medical} & CT   & 2  & 63    & Lung nodule \\\hline
    BCV   \cite{bcv} & CT      & 14          & 30    & Abdominal organs\\\hline
    AMOS CT \cite{ji2022amos} & CT    & 16      & 200    & Abdominal organs\\\hline
    AMOS MR \cite{ji2022amos} & MRI   & 16      & 40   & Abdominal organs \\      
    \bottomrule
    \end{tabular}
    \end{threeparttable}
    \label{tab:dataset_details}}
\end{table}

We further evaluate the models' generalization on seven public 3D medical image segmentation datasets, as shown in Fig. \ref{fig:cover_fig} and Tab. \ref{tab:pub}. To ensure unbiased and fair comparisons, we re-implement all models and assess their performance using five-fold cross-validation on the available training set unless other noted, rather than relying on the official testing platforms, enabling objective and equitable model evaluation under a consistent framework. Notably, our experiments avoid additional performance-boosting techniques common in challenges for these datasets, such as model ensembles, test-time augmentation, or post-processing. As a result, we focus on assessing the core-model capabilities under the same evaluation strategy.

See in Table \ref{tab:dataset_details}, the selected datasets cover diverse scenarios in medical imaging, covering a range of target types such as healthy tissues, organs, diseased organs, and tumors; dataset sizes, from as small as 30 samples to the relatively large size of 210 samples; and various image modalities. Note the performance of AMOS CT and MR are reported on the official validation set, and the results of UNETR, nnFormer and SwinUNETR are cited from the AMOS benchmark paper \cite{ji2022amos}. The results of nnUNet in Table \ref{tab:pub} are cited from \cite{isensee2021nnu}, and all others are obtained with our framework under cross-validation.

Fig. \ref{fig:cover_fig} demonstrates that MedFormer consistently outperforms all other comparison methods across seven datasets. Interestingly, the recently proposed Transformer based models are not as generalizable as nnUNet. Although they perform well on some specific tasks, e.g. nnFormer on the liver tumor, they usually perform worse on most datasets. Table \ref{tab:pub} shows that our framework-trained ResUNet has comparable performance to nnUNet, while attention modules or atrous spatial pyramid pooling offer slight improvements in Attention UNet and DeepLabV3+. 

UNETR underperforms across all datasets, indicating the ViT encoder is unsuitable for dense prediction tasks due to the loss of structural information within image patches. Models with Swin attention, e.g. nnFormer, SwinUNETR and VT-UNet, perform better but still lag behind ConvNets due to limited medical image training data. Notably, VT-UNet uses ImageNet pretraining weights; however, transfer learning is not effective due to the significant domain gap between 3D medical images and natural images.

MedFormer's hybrid architecture leverages the advantages of both ConvNet and Transformer, achieving the best performance with a large margin even with limited data (BCV and AMOS MR) or small and low-contrast regions of interest (liver tumor, kidney tumor, and lung nodule).

\begin{figure*}[]
    \centering
    \includegraphics[width=\textwidth]{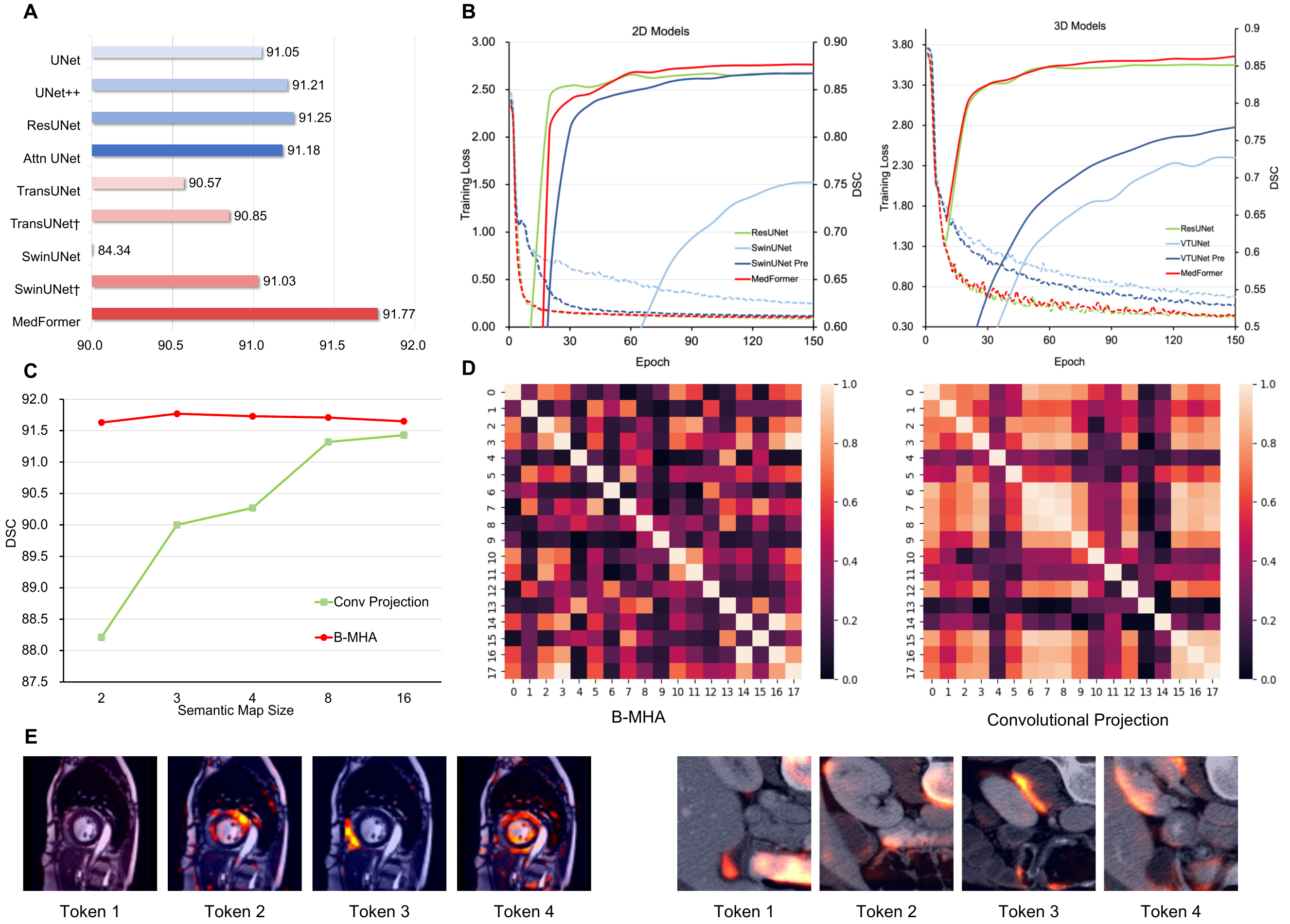}
    \caption{(A): DSC of 2D models on the ACDC dataset. (B): Convergence analysis. The testing Dice score and training loss v.s. training epochs on 5\% data of the collected large cardiac dataset in 2D setting, and on BCV dataset in 3D setting. Solid line: Dice score. Dashed line: training loss. (C): Comparison of the semantic map size between B-MHA and linear convolutional projection on the ACDC dataset. (D): The cosine similarity matrices of tokens in $2\times 3\times3$ semantic map projected by B-MHA and convolutional projection on the BCV dataset in the 3D setting. (E) Attention map visualization, each column represents the attention map of one selected semantic token on the full image token map. Left: One slice of a cardiac image from the ACDC dataset. Right: One image patch of a CT image from the BCV dataset.}
    \label{fig:fig4}
\end{figure*}

\subsection{Convergence analysis.}
Fig. \ref{fig:fig4} (B) displays the training loss and test DSC curves in both 2D and 3D settings. We observe that the introduced inductive bias allows MedFormer to achieve rapid convergence similar to CNNs, while offering higher capacity. In the 2D setting, using 5\% of the collected cardiac dataset for training, MedFormer demonstrates convergence speed comparable to ResUNet, but with superior performance. In contrast, SwinUNet with a pure Transformer architecture converges considerably slower without ImageNet pretraining weights. When initialized with ImageNet pretrained weights, SwinUNet's learning speed increases but still falls short of ResUNet or MedFormer.

Training 3D models poses greater challenges than their 2D counterparts. When utilizing the BCV dataset for training, MedFormer achieves rapid convergence comparable to ResUNet. However, VT-UNet, featuring a pure Transformer architecture, converges considerably slower due to optimization challenges arising from the absence of inductive bias and limited training data. Even when initialized with ImageNet pretrained weights, VT-UNet's convergence remains slow, as the domain gap between 3D medical images and natural images is more pronounced than in 2D settings. Transfer learning from ImageNet offers minimal benefit for 3D medical image analysis. MedFormer, with its specially designed architecture, can learn effectively from scratch without requiring pretraining.

\subsection{Token compression analysis on B-MHA.}

The proposed B-MHA effectively reduces the quadratic computational complexity of traditional self-attention to linear by non-linearly projecting high-resolution token maps to a low-rank representation. This design facilitates the retention of essential semantics while significantly minimizing redundancy. To examine the influence of semantic map size on performance, we conducted ablation studies using the ACDC dataset. As illustrated in Fig. \ref{fig:fig4} (C), segmentation accuracy substantially decreases as semantic map size diminishes when attention with convolutional projection is employed. This reduction in accuracy can be ascribed to the ineffectiveness of linear compression in retaining valid tokens, resulting in information loss and compromised performance. In contrast, B-MHA exhibits high performance even when the semantic map size is as small as $2\times 2$.

To further validate our hypothesis that B-MHA has a better compression ability to reduce redundancy, we computed the cosine similarity among the compressed tokens in the semantic map produced by B-MHA and convolutional projection. We set the size of the semantic map to $2\times 3\times 3$ for the BCV dataset. The heatmap of the cosine similarity matrix is presented in Fig. \ref{fig:fig4} (D), where absolute values are used for ease of visualization. The block-like pattern observed in the convolutional projection suggests that some adjacent tokens share similar semantics. In contrast, the cosine similarity of tokens compressed by B-MHA is considerably lower, indicating increased orthogonality and reduced redundancy.

\subsection{Attention map visualization.}

We further investigate the learned features of B-MHA by visualizing the attention map of the tokens in the semantic map, as displayed in Fig. \ref{fig:fig4} (E). In this visualization, tokens in the semantic map aggregate semantic information from the full image token map, using the attention map as weights; brighter regions means higher weights for aggregation. Fig. \ref{fig:fig4} (E) (left) presents the attention on the ACDC dataset in a 2D setting. Token 1 concentrates on the background, token 2 corresponds to MYO, token 3 is associated with RV, and token 4 focuses on LV. Fig. \ref{fig:fig4} (E) (right) demonstrates the attention of a $32\times 128 \times 128$ image patch from the BCV dataset in a 3D setting. Token 1 attends to the stomach, token 2 emphasizes the intervertebral regions, and tokens 3 and 4 highlight the outer and inner boundaries of the kidney, respectively.

Without explicit supervision on the semantic map, MedFormer automatically learns class-specific features by globally aggregating information according to the attention weights.This observation further substantiates the ability of B-MHA to effectively eliminate redundancy in image tokens and compress them into semantically meaningful semantic tokens. Through visualization experiments, we have demonstrated that these tokens exhibit remarkable interpretability.

\section{Discussion}

We offer key insights into model pre-training, architecture design for 3D medical images. Also, we present technical recommendations towards building robust medical image segmentation workflows.

\textbf{Pre-training.} Although Transformers have exhibited superior abilities across visual tasks, the search for proper training strategies for medical applications remains unsolved. Large-scale pre-training is an indispensable requirement for standard Transformers to work well in downstream tasks in the NLP\cite{devlin2018bert} and CV\cite{dosovitskiy2020image} fields. However, our findings indicate that transferring pre-trained weights from natural images to the medical field can be sub-optimal due to the significant distribution gap. One key difference between 2D natural images and 3D medical images is that natural images typically contain a single primary object, making contrastive based or masked image modeling based self-supervised learning effective. In contrast, 3D medical images often include multiple organs and diverse tissues within a single image. Consequently, embedding an entire image into a token for contrastive learning or reconstructing one organ from its neighboring organs may not be as effective. Directly applying self-supervised pre-training designed for natural images may not yield satisfactory results for 3D medical tasks, and the development of suitable pre-training methods for 3D medical images remains an open problem. To address these challenges, MedFormer greatly alleviates the data demand while maintaining model capacity. In particular, the data-scalable property frees MedFormer from complex pre-training task design, allowing it to analyze a wide range of medical image tasks.

\textbf{The design of 3D medical imaging architecture.} After systematically comparing the performance of Transformers and CNN-based models in a broad range of medical image segmentation tasks, we found that previously proposed Transformer models do not consistently outperform CNN models. The reasons behind this finding are multifaceted. Although Transformers have a greater capacity for modeling complex structures, ViT-like Transformers face computational complexity constraints that lead to excessive downsampling of the input, rendering them unsuitable for dense prediction tasks such as segmentation. Swin-based Transformers, lacking inductive bias, struggle to effectively train on limited medical image data. Furthermore, we verified that large-scale pre-trained weights on natural images do not transfer well to medical images, especially in 3D settings. On the other hand, CNNs excel at capturing local textures and can learn with a small amount of data due to their inherent inductive bias. As a result, hybrid architectures like MedFormer successfully combine the strengths of both approaches, leveraging their advantages and demonstrating improved segmentation performance across various medical imaging scenarios.

\textbf{Recommendations for building generalizable medical model.} Our findings offer several key insights into model design, evaluation, and generalization for diverse segmentation tasks. First, it is essential to carefully assess the individual contributions of the Transformer and CNN components in hybrid models. For example, although TransUNet introduces ViT to the ResNet backbone, adding numerous parameters, its performance is comparable to R50-UNet. Second, establishing solid core-model baselines is crucial for unbiased model evaluations. Reporting results for Transformers with auxiliary techniques (e.g., data augmentation, advanced optimizers, or model ensembles) does not faithfully assess the fairness of core architectures. Our results, obtained through a consistent evaluation framework, indicate that the core-architecture performance of previous Transformer models does not consistently outperform CNN baselines. Third, evaluating task-agnostic datasets is vital for measuring the generalization ability of segmentation models. Current Transformer-related studies are highly task-dependent, whereas CNN-based U-Net models have been verified across various tasks \cite{isensee2021nnu}. Our study highlights the importance of using multi-dataset evaluations with diverse anatomies for assessing generalization. Additionally, incorporating datasets from multi-center, multi-vendor, or different scanning protocols is crucial for a robust evaluation against medical domain shifts.

\textbf{Limitation and future work.}
Recent advancements in large-scale models in NLP \cite{brown2020language,touvron2023llama} and CV \cite{radford2021learning,li2022blip,li2022grounded} fields demonstrate the impact of massive, multi-task, and multi-modal training on improving model performance, robustness, and generalization. However, in the field of medical image analysis, the prevailing training paradigm remains focused on training separate models for specific medical tasks or datasets. This approach has limitations, as it cannot effectively utilize available medical images and fails to provide the model with a comprehensive understanding of human physiological structures during training. As a result, we plan to investigate the utility of large-scale foundation models for medical imaging that can learn through a multi-task, multi-modal, and multi-body training paradigm. Furthermore, our current work is limited to supervised training, and exploring the integration of self-supervised or semi-supervised training is a valuable direction for future research. In addition, an effective model should possess strong transfer learning and few-shot learning capabilities. Finally, we will explore human-in-the-loop learning, refining the model based on doctors' feedback on model's prediction \cite{diao2021efficient}. This approach will help to improve the model's performance while ensuring that the model's prediction is consistent with medical professionals in real-world settings.

\section{Conclusion}
In this study, we present MedFormer, a hybrid Transformer segmentation model that is scalable across data amounts, ranging from small-scale to large-scale data without pre-training. MedFormer showcases its potential for generalization by establishing new state-of-the-art baselines across seven widely-used datasets with different image modalities (e.g., CT and MRI) and target anatomies (e.g., healthy organs, diseased tissue, and tumors).  The strong performance of MedFormer can be attributed to the desirable inductive bias introduced by the unified architecture and the efficient B-MHA module, which learns global semantics through low-rank projection. These key designs enable MedFormer to capture boundary details, fuse global information in a hierarchical manner, and exhibit robustness against data distribution shifts. Furthermore, we provide a comprehensive codebase for fair comparison of different architectures across a wide range of medical image segmentation tasks. We expect that this codebase will serve as a solid baseline for future medical image model design, implementation, and evaluation, thereby driving advancements in the field.

\bibliographystyle{IEEEtran}
\bibliography{reference.bib}
%



\end{document}


%
\title{A Data-scalable Transformer for Medical Image Segmentation: Architecture, Model Efficiency, and Benchmark}
%
%
%
%


\author{Yunhe~Gao,
        Mu~Zhou,
        Di~Liu,
        Zhennan~Yan,
        Shaoting~Zhang,
        and Dimitris~N.~Metaxas
\IEEEcompsocitemizethanks{\IEEEcompsocthanksitem Y. Gao, D. Liu and D. Metaxas are with the Computer Science Department, Rutgers University, Piscataway,
NJ, USA.\protect\\
\IEEEcompsocthanksitem M. Zhou and Z. Yan are with SenseBrain Research, Princeton, NJ, USA.\protect
\IEEEcompsocthanksitem S. Zhang is with hanghai Artificial Intelligence Laboratory, Shanghai, China.}
\thanks{Manuscript received October 15, 2022.}}

\IEEEdisplaynontitleabstractindextext

%
\IEEEpeerreviewmaketitle

\appendices
\section{Datasets details} \label{appendix:datasets}

\subsection{UK Biobank dataset.}
The UK Biobank (UKBB) dataset\cite{petersen2015uk,petersen2017reference} comprises 1,008 short-axis cardiac MR images collected from multiple centers using standardized protocols. The dataset provides annotations for three categories at the end-diastolic (ED) and end-systolic (ES) phases: left ventricle (LV), right ventricle (RV), and left ventricular myocardium (MYO). The images were acquired using a clinical wide-bore 1.5 Tesla scanner from Siemens Healthcare. The UKBB is a population-based prospective study designed to facilitate comprehensive investigations of genetic and environmental factors influencing diseases. The cohort includes 500,000 voluntary participants, aged between 40 and 69 years, who were recruited between 2006 and 2010 across the UK. Starting in 2014, 100,000 volunteers from the entire cohort also participated in multi-modal imaging, encompassing MR imaging of the brain, heart, and whole body. In this work, we use 1,008 short-axis cardiac MR images as training data. It is important to note that, as the UKBB is a prospective study, these patients do not necessarily have heart disease.

\subsection{ACDC dataset.}

The ACDC dataset\cite{bernard2018deep} (Automated Cardiac Diagnosis Challenge) comprises cardiac short-axis Cine MRI data from 100 patients, collected at the University Hospital of Dijon, France. Annotations are provided for three categories at the end-diastolic (ED) and end-systolic (ES) phases, including left ventricle (LV), right ventricle (RV), and left ventricular myocardium (MYO). The dataset encompasses healthy patients, as well as those with previous myocardial infarction, dilated cardiomyopathy, hypertrophic cardiomyopathy, and abnormal right ventricle, with 20 scans for each group. These data were acquired over a 6-year period using two MRI scanners with two magnetic strengths (1.5T and 3.0T, Siemens Healthcare). The dataset includes a series of short-axis slices that cover the LV from the base to the apex, with a thickness ranging between 5-10 mm. Spatial resolution varies from 0.70 to 1.92 $mm^2$/pixel, while 28 to 40 images fully or partially cover the cardiac cycle.

\subsection{M\&Ms and M\&Ms-2 dataset.}

The M\&Ms (Multi-centre, multi-vendor, and multi-disease cardiac segmentation) and M\&Ms-2 (Multi-Disease, Multi-View \& Multi-Center Right Ventricular Segmentation in Cardiac MRI)\cite{campello2021multi} challenges are organized in conjunction with MICCAI and STACOM. The M\&Ms challenge cohort comprises patients with hypertrophic and dilated cardiomyopathies, as well as healthy subjects. The subjects were scanned at clinical centers in three countries (Spain, Germany, and Canada) using four different MRI scanner vendors (Siemens, General Electric, Philips, and Canon). The training set contains 150 annotated images from two vendors (75 each), while the testing set consists of 170 cases (20 for the first vendor and 50 each for the other three vendors). The M\&Ms-2 cohort was collected at three clinical centers in Spain using three different magnetic resonance scanner vendors (Siemens, General Electric, and Philips). The training set comprises 160 annotated images, including 40 normal subjects, 30 dilated left ventricle subjects, 30 hypertrophic cardiomyopathy subjects, 20 congenital arrhythmogenesis subjects, 20 tetralogy of fallot subjects, and 20 interatrial communication subjects. Long-axis MR images are also provided in the M\&Ms-2 dataset, but they are not used in this study. Annotations are available for three categories at the end-diastolic (ED) and end-systolic (ES) phases, including left ventricle (LV), right ventricle (RV), and left ventricular myocardium (MYO).

\subsection{BCV dataset.}

The BCV dataset\cite{bcv} (Multi-Atlas Labeling Beyond the Cranial Vault) comprises 50 subjects with abdominal CT scans, of which only 30 training images are publicly available. We conduct experiments using the 30 training CT scans for 5-fold cross-validation. Thirteen abdominal organs were manually labeled by two experienced undergraduate students and verified by a radiologist on a volumetric basis using the MIPAV software. The labeled organs include the spleen, right kidney, left kidney, gallbladder, esophagus, liver, stomach, aorta, inferior vena cava, portal vein and splenic vein, pancreas, right adrenal gland, and left adrenal gland. Some patients may lack the right kidney or gallbladder, and therefore these organs are not labeled. All scans were acquired for routine clinical care from CT scanners at the Vanderbilt University Medical Center (VUMC). Trained raters manually labeled all datasets, and a radiologist or radiation oncologist reviewed the labels for accuracy. The scans were captured during the portal venous contrast phase with variable volume sizes ($512\times 512 \times 85$ - $512 \times 512 \times 198$) and field of views (approximately $280\times 280 \times 280 mm^3$ - $500 \times 500 \times 650 mm^3$). The in-plane resolution ranges from $0.54 \times 0.54 mm^2$ to $0.98 \times 0.98 mm^2$, while the slice thickness varies from $2.5 mm$ to $5.0 mm$.

\subsection{LiTS dataset.} 
The LiTS dataset\cite{bilic2019liver} (Liver Tumor Segmentation Challenge) comprises 201 computed tomography (CT) images of the abdomen, with 131 training cases and 70 testing cases. Among these, 194 CT scans include liver lesions. We conduct experiments using the 131 training dataset for 5-fold cross-validation. The LiTS dataset provides detailed annotation for tumors while offering coarse annotation for the liver. The image data originates from various clinical sites, including Ludwig Maxmilian University of Munich, Radboud University Medical Center of Nijmegen, Polytechnique \& CHUM Research Center Montréal, Tel Aviv University, Sheba Medical Center, IRCAD Institute Strasbourg, and the Hebrew University of Jerusalem. The studied subjects suffer from diverse liver tumor diseases, such as hepatocellular carcinoma (HCC), as well as secondary liver tumors and metastases originating from colorectal, breast, and lung cancers. The tumors exhibit varying contrast enhancement, including hyper and hypo-dense contrast. The images represent a mix of pre- and post-therapy abdominal CT scans, acquired with different CT scanners and acquisition protocols. The image data is diverse in terms of resolution and quality. Image resolution ranges from $0.56\ mm$ to $1.0\ mm$ in the axial direction and $0.45\ mm$ to $6.0\ mm$ in the z-direction. The number of slices in the z-direction varies from $42$ to $1026$. Some images contain imaging artifacts (e.g., metal artifacts) commonly found in real-life clinical data. The number of tumors ranges from $0$ to $75$, with tumor sizes varying between $38\ mm^3$ and $349\ cm^3$.

\subsection{KiTS dataset.} 
The KiTS19 dataset\cite{heller2019kits19} comprises segmented CT imaging and treatment outcomes for 300 patients who underwent partial or radical nephrectomy for one or more kidney tumors at the University of Minnesota Medical Center between 2010 and 2018. Out of these cases, 210 have been released publicly, while the remaining 90 are kept private for evaluation purposes. The challenge aims to accelerate the development of reliable kidney and kidney tumor semantic segmentation methodologies. We conduct experiments using the 210 publicly available training images for 5-fold cross-validation.

\subsection{MSD Lung Nodule dataset}
The MSD Lung dataset is a part of the Medical Image Segmentation Decathlon (MSD) \cite{antonelli2022medical}, an international challenge aimed at identifying a general-purpose algorithm for medical image segmentation. The competition encompasses ten distinct datasets featuring various target regions, modalities, and challenging attributes. We chose the MSD lung nodule segmentation dataset to assess the models' performance on small and randomly distributed targets. We conduct a 5-fold cross-validation on the publicly available training set, consisting of 63 CT scans annotated with lung nodules.

\subsection{AMOS dataset}
The AMOS dataset \cite{ji2022amos} is a large-scale collection of CT and MRI data from 600 patients diagnosed with abdominal tumors or abnormalities at Longgang District People's Hospital. The dataset comprises 500 CT and 100 MRI scans acquired from eight different scanners and vendors, encompassing 15 organ categories: spleen, right kidney, left kidney, gallbladder, esophagus, liver, stomach, aorta, inferior vena cava, pancreas, right adrenal gland, left adrenal gland, duodenum, bladder, and prostate/uterus. For CT images, AMOS provides 200 scans for training and 100 scans for validation, while for MRI images, 40 scans are designated for training and 20 scans for validation. In line with the AMOS benchmark paper \cite{ji2022amos}, we report performance on the validation set and utilize all training data for model training.

\section{Comparison methods}\label{appendix:models}

In this study, we systematically compare the proposed MedFormer with multiple CNN-based and Transformer-based segmentation models in both 2D and 3D settings, including:

\begin{itemize}
    \item \textbf{UNet} \cite{ronneberger2015u} is a well-known medical image segmentation model. UNet has an encoder-decoder architecture, consisting of a contracting path to capture context and a symmetric expanding path that enables precise localization. We implement the original UNet in both 2D and 3D settings, which employ double convolution layers on every level.
    \item \textbf{UNet++} \cite{zhou2018unet++}is a UNet variant introducing nested, dense skip connections in the encoder and decoder sub-networks to reduce the semantic gap between feature maps, resulting in improved feature fusion.
    \item \textbf{Attention UNet} \cite{oktay2018attention} adds an attention gating module to UNet, which learns to suppress irrelevant regions and emphasize salient features relevant to a specific task. We implement the original Attention UNet in 2D and modify the 3D setting by adding attention gating to ResUNet for enhanced gradient flow, faster convergence, and better performance.
    
    \item \textbf{ResUNet} s a UNet variant that replaces double convolution layers with residual blocks from ResNet \cite{he2016identity}. Residual connections facilitate faster convergence and improved performance, implemented in both 2D and 3D settings.
    
    \item \textbf{DeepLabV3+} \cite{chen2017deeplab} is a state-of-the-art semantic segmentation algorithm that builds upon the success of its predecessor, DeepLabV3. It aims to provide accurate and efficient segmentation of objects and regions in images by combining the strengths of deep convolutional neural networks (CNNs) and atrous spatial pyramid pooling (ASPP) modules. DeepLabV3+ introduces a novel encoder-decoder architecture, which enhances the model's ability to capture fine-grained details and recover object boundaries more effectively. The encoder is responsible for extracting high-level semantic features, while the decoder refines the segmentation results by incorporating low-level features. This combination results in improved segmentation performance across various object scales and complexities. We implement DeepLabV3+ in 3D based on the ResUNet encoder-decoder architecture. 

    \item \textbf{nnUNet} \cite{isensee2021nnu} is an automated segmentation framework for medical images. It utilizes UNet as its backbone, but offers a very specialized training technique and hyper-parameter configuration. nnUNet achieves state-of-the-art performance on several medical image segmentation challenges without a complex architectural design. In the three public dataset experiments, we cite the nnUNet paper's results as a solid CNN baseline.

    \item \textbf{TransUNet} \cite{chen2021transunet} is an early attempt to Transformers on medical image segmentation. It uses a R50-ViT \cite{dosovitskiy2020image} as the encoder and employs a UNet decoder for 2D medical image segmentation. Specificially, the images is first fed to the ResNet50 backbone\cite{he2016identity} to obtain the feature map ($16\times$ down-sampled), and then processed by ViT for further modeling. 
    \item \textbf{R50-UNet} \cite{he2016deep} is used as an ablation with TransUNet. It uses ResNet-50 as the encoder while the decoder is the same with TransUNet. The only difference is that ResNet50-UNet doesn't have ViT on top of the ResNet-50 backbone.
    
    \item \textbf{SwinUNet} \cite{cao2021swin} is a pure Transformer model for 2D medical image segmentation. It uses SwinTransformer blocks\cite{liu2021swin} in a UNet-like encoder-decoder structure.  It first divides images into $4\times 4$ patches and then projects patches into token embeddings. The tokens are processed by interleaved window-based attention and shifted window-based attention modules. The window-based attention computes the attention locally and propagates information across windows by shifting the window.

    \item \textbf{UNETR} \cite{hatamizadeh2021unetr} uses a ViT-like structure as the encoder and uses a CNN decoder for 3D medical image segmentation. It first divides 3D images into patches, and then linearly projects the patches into token embeddings. The tokens are then processed by the self-attention block, similar to ViT. Due to the quadratic complexity of conventional self-attention, the sequence length can't be too long. Thus the patch size is large (e.g. $16\times16\time16$).

    \item \textbf{SwinUNETR} \cite{hatamizadeh2022swin} is similar to UNETR but replaces the ViT encoder with SwinTransformer-based encoder.

    \item \textbf{nnFormer} \cite{zhou2021nnformer} is a 3D transformer for volumetric
    medical image segmentation. nnFormer uses an interleaved combination of convolution and self-attention operations. It also utilizes both local and global volume-based self-attention to build feature pyramids and provide large receptive fields. Skip attention is proposed to replace traditional concatenation/summation operations in skip connections
    
    \item \textbf{VT-UNet} \cite{peiris2021volumetric} is a pure Transformer model for 3D medical image segmentation. It adapts the SwinTransformer block to 3D in both the encoder and decoder. Similar to SwinUNet, VT-UNet first linearly projects patches into token embeddings and then processes tokens by interleaved 3D window-based attention module and 3D shifted window-based attention module in the encoder and decoder.
    
    \item \textbf{UTNet} \cite{gao2021utnet} is a hybrid model that employs interleaved convolution and Transformer blocks for 2D medical image segmentation. To extract long-range relationships on high-resolution token maps, unlike the local attention of SwinTransformer or excessive down-sampling as in ViT, UTNet uses efficient attention that reduces the token length by eliminating redundancy in the token sequences.
\end{itemize}

\section{Experiment details}\label{appendix:experiments}

\subsection{Experiments setup}

Our experiments can be divided into two groups based on their objectives. The first set of experiments aims to determine the effect of training data scale on models. We collect a large short-axis cardiac cine MRI dataset with the same data modality and annotations, including UKBB, ACDC, M\&Ms, and M\&Ms-2. We employ the M\&Ms testing set as our testing set, while the remaining datasets serve as our training set. As M\&Ms testing set is sufficiently large (170 cases, 340 MR scans on ED and ES phases) and contains four vendor domains (A: Siemens, B: Philips, C: General Electric, D: Canon), we use it to evaluate models' performance and robustness in lieu of cross-validation. It should be noted that images from UKBB and ACDC are scanned with Siemens scanners, while images from MnMs-2 are scanned with Siemens, Philips, and General Electric. Therefore, our training set is dominated by images from domains A and B, with only a small amount of images from domain C and no image from domain D. We identify vendor C as a rarely seen domain and vendor D as an unseen domain for robustness evaluation. In addition, the second group of experiments evaluates the generalization capability of models on diverse medical datasets, including ACDC, BCV, and LiTS. These experiments cover a variety of medical contexts, including MRI and CT scans; healthy organs, diseased organs, and malignant tumors; 2D and 3D settings. Instead of utilizing the official online evaluation site, we conduct five-fold cross-validation on the collected training samples of these datasets. We underline the importance of this setting for a fair evaluation and accurate performance comparison. Post-processing, model ensemble, and test-time augmentation are common performance-enhancing refinement techniques utilized in the challenge submissions on these datasets. These techniques are model-agnostic and obscure the models' actual performance. In contrast, we train all models within our framework using the same training methodologies and evaluate the metrics immediately based on the models' output without any refining methods.

\subsection{Implementation details}
Our framework are implemented using PyTorch framework and all models are trained on one NVIDIA A100 GPU with 40 GB memory. AdamW optimizer with an exponential learning rate decay is used for optimization. Data augmentations are applied on the fly during training, including random rotation, scaling, brightness shift, contrast adjustment, Gaussian noise, and random crop. We use cross-entropy loss:
\begin{equation}
    \mathcal{L}_{CE}=\frac{1}{C}\sum_{t=1}^{C}y_t\log(\hat{y}_t)
\end{equation}

and Dice loss:
\begin{equation}
    \mathcal{L}_{DL}=\frac{1}{C}\sum_{t=1}^{C}(1-\frac{2\sum_{i\in I}y^i_t\hat{y}_t^i}{\sum_{i\in I}y_t^i + \sum_{i\in I}\hat{y}_t^i})
\end{equation}
for training, where $C$ is the number of classes, $I$ indicates the whole image, $y_t$ and $\hat{y}_t$ are the ground truth and the predicted segmentation from the model of class $t$, respectively. The total loss is the summation of cross-entropy loss and Dice loss:
\begin{equation}
    \mathcal{L}_{total} = \mathcal{L}_{CE} + \mathcal{L}_{DL}
\end{equation}

For data preprocessing, we follow the setting of nnUNet \cite{isensee2021nnu}. For 2D settings, the in-plane of all images is resampled to the median spacing of the entire dataset, while the through-plane spacing remains unmodified. Images are resampled to the median spacing along all three axes for 3D settings. Images are then normalized according to the intensity of the foreground. During evaluation for 2D models, 3D images are predicted slice by slice and then stacked together as the final prediction. For 3D models, images are predicted with a sliding window approach, where the window size equals the patch size used during training. Adjacent predictions overlap by half the size of a patch. The final prediction is an average of all predictions (after softmax) on the overlapped regions. Due to the diversity of data, tasks, and settings involved in this paper, the additional experimental details for each dataset can be found in the following or in our codebase\footnote{https://github.com/yhygao/CBIM-Medical-Image-Segmentation}.

The large cardiac MRI dataset, a combination of UKBB, ACDC, M\&Ms, and M\&Ms-2, is used with 2D settings for all models. Images are resampled to an in-plane spacing of $1.25\times 1.25\ mm$ and through-plane spacing remains unchanged. Training data is randomly shuffled and partitioned into various scales. All models are trained for 150 epochs with a batch size of 32. In MedFormer, the number of B-MHA blocks for each level is $N_1=2, N_2=2, N_3=2$, and the semantic map size is set to $4\times 4$.

For the ACDC dataset, both 2D and 3D settings are employed. In the 2D setting, images are resampled to an in-plane spacing of $1.56\times 1.56$, while in the 3D setting, images are resampled to $1.56\times 1.56\times 5.0\ mm$. Data is randomly shuffled and split into five folds for cross-validation. 2D models are trained for 200 epochs with a batch size of 32 and an input size of $256\times 256$. 3D models are trained for 200 epochs with 200 iterations per epoch (bootstrap sampling) and an input size of $16\times 192\times 192$. In MedFormer, the number of B-MHA blocks for each level is $N_1=2, N_2=2, N_3=2$, and the semantic map size is set to $3\times 3$ for 2D and $3\times 3 \times 3$ for 3D.

For the BCV dataset, a 3D setting is used for all models, and images are resampled to $0.75\times 0.75\times 3.0\ mm$. Data is randomly shuffled and split into five folds for cross-validation. UNETR and VT-UNet are trained for 300 epochs with 300 iterations per epoch, while other models are trained for 150 epochs with 300 iterations per epoch. The batch size is set to 3, and MedFormer's input size is $32\times 128\times 128$. The number of B-MHA blocks in each level of MedFormer is $N_1=2, N_2=4, N_3=6$, and the semantic map size is set to $4\times 4\times 4$.

For the LiTS dataset, a 3D setting is used for all models, and images are resampled to $0.76\times 0.76\times 1.0\ mm$. UNETR and VT-UNet are trained for 400 epochs with 500 iterations per epoch, while other models are trained for 200 epochs with 500 iterations per epoch. The batch size is set to 2, and MedFormer's input size is $128\times 128\times 128$. The number of B-MHA blocks in each level of MedFormer is $N_1=2, N_2=4, N_3=6$, and the semantic map size is set to $4\times 4\times 4$.

For the KiTS dataset, a 3D setting is used for all models, and images are resampled to $0.78\times 0.78\times 0.78\ mm$. UNETR and VT-UNet are trained for 400 epochs with 500 iterations per epoch, while other models are trained for 300 epochs with 500 iterations per epoch. The batch size is set to 2, and MedFormer's input size is $128\times 128\times 128$. The number of B-MHA blocks in each level of MedFormer is $N_1=2, N_2=4, N_3=6$, and the semantic map size is set to $4\times 4\times 4$.

For the AMOS dataset, a 3D setting is used for all models. CT images are resampled to $0.68\times 0.68\times 2.0\ mm$ and MR images to $1.19\times 1.19\times 2.0\ mm$. The official training set is used for training, and performance is reported on the official validation set. Results for nnUNet, UNETR, SwinUNETR, and nnFormer are cited from \cite{ji2022amos}, while other models' results are obtained using our framework. For CT modality, models are trained for 400 epochs with 500 iterations per epoch; for MR modality, models are trained for 200 epochs with 80 iterations per epoch. The batch size is set to 2, and MedFormer's input size is $128\times 128\times 128$. The number of B-MHA blocks in each level of MedFormer is $N_1=2, N_2=4, N_3=6$, and the semantic map size is set to $4\times 4\times 4$.

For the MSD Lung Nodule dataset, a 3D setting is used for all models, and images are resampled to $0.78\times 0.78\times 1.24\ mm$. All models are trained for 300 epochs with 500 iterations per epoch, and the batch size is set to 2. To stabilize the training process, at least one image patch in the batch must contain a lung nodule. Specifically, the first image in the training batch is cropped around the lung nodule, and the second image in the batch undergoes random cropping. MedFormer takes $128\times 128\times 128$ image patches as input. The number of B-MHA blocks used in each level of MedFormer is $N_1=2, N_2=4, N_3=6$, and the semantic map size is set to $4\times 4\times 4$.

\subsection{Evaluation metrics}

We use two types of metrics to evaluate all models. The Dice similarity coefficient (DSC) evaluates the overlap of the predicted and ground truth segmentation map:
\begin{equation}
    DSC = \frac{2|P\cap G|}{|P| + |G|},
\end{equation}
where $P$ indicates the predicted segmentation map and $G$ denotes the ground truth. A DSC of 1 indicates a perfect segmentation while 0 indicates no overlap at all.

Hausdorff distance (HD) measures the largest symmetrical distance between two segmentation maps:
\begin{equation}
    d_{H}(P, G) = \max\left\{ \sup\limits_{p \in P} \inf\limits_{g\in G} d(p, g), \sup\limits_{g \in G} \inf\limits_{p\in P} d(p, g)  \right\},
\end{equation}
where $d(\cdot)$ represents the Euclidean distance, $\sup$ and $\inf$ denote supremum and infimum, respectively. We employ 95\% HD to eliminate the impact of a very small subset of the outliers.

\section{Supplementary results}

\subsection{Complexity analysis.}
The computation complexity comparison is shown in Table \ref{tab:complexity}. Given a feature representation map with size $d\times H\times W$ and a semantic map with size $d\times h\times w$, where $h, w \ll H, W$ are fixed values. The Convolutional layers, shifted-window (Swin) self-attention and the proposed B-MHA have linear complexity to the input length $HW$, while the vanilla self-attention has quadratic complexity that is unaffordable for high-resolution feature maps. Note the SwinTransformer needs to consecutively stack a window-based self-attention and a shifted-window self-attention to model connections across windows, so the complexity and parameters are doubled.

\begin{table*}[h]
    \centering

    \caption{The computation complexity comparison.}
    \begin{threeparttable}
    \begin{tabular}{c|c|c}
    \toprule
        Module                & Complexity                                          & parameters\\ \hline
        Conv            & $\Omega(k^2HWd^2)$                                        & $k^2d^2$\\ \hline
        MHSA             & $\Omega(4HWd^2+2(HW)^2d)$                                 & $4d^2$  \\ \hline
        W-MHA+SW-MHA \cite{liu2021swin}    & $\Omega(8HWd^2 + 4M^2HWd))$    & $8d^2$\\ \hline
        B-MHA           &     $\Omega(3HWd^2+3hwHWd)$           & $6d^2$   \\
    \bottomrule

    \end{tabular}
    
    \begin{tablenotes}
    \footnotesize
        \item $k$: convolution kernel size, $HW$: the input token number, $d$: token dimension, $M$: the window size of shifted-window (Swin) self-attention, $hw$: the semantic map token number. \\
    \end{tablenotes}
    \end{threeparttable}
    \label{tab:complexity}
\end{table*}

\subsection{Inference time and GPU memory consumption}
Besides theoretical complexity analysis, we provide real inference time and GPU memory consumption of the comparison 3D models. We use the image tensor size of $2\times 1\times64\times 128\times 128$ (batch size 2) as the input. We report the average of 500 runs.

\begin{table*}[h]
    \centering

    \caption{The comparison of inference time and GPU memory consumption}
    \begin{threeparttable}
    \begin{tabular}{c|c|c|c|c|c|c|c}
    \toprule
                    & ResUNet & Attn UNet & UNETR & SwinUNETR & nnFormer & VT-UNet & MedFormer \\\midrule
     Inference Time/s &0.083	&0.094	&0.035	&0.118 &	0.067	&0.121 & 0.107\\\hline
     Memory/GB & 19.52	&21.99&	7.12	&29.87&	15.06	&19.29&	24.66 \\
    \bottomrule

    \end{tabular}
    
    \begin{tablenotes}
    \footnotesize
        \item Measured with tensor size of $2\times 1\times64\times 128\times 128$ (batch size 2) \\
    \end{tablenotes}
    \end{threeparttable}
    \label{tab:complexity}
\end{table*}

\subsection{Supplementary ablation Study}

\begin{figure*}[h]
    \centering
    \includegraphics[width=1.0\textwidth]{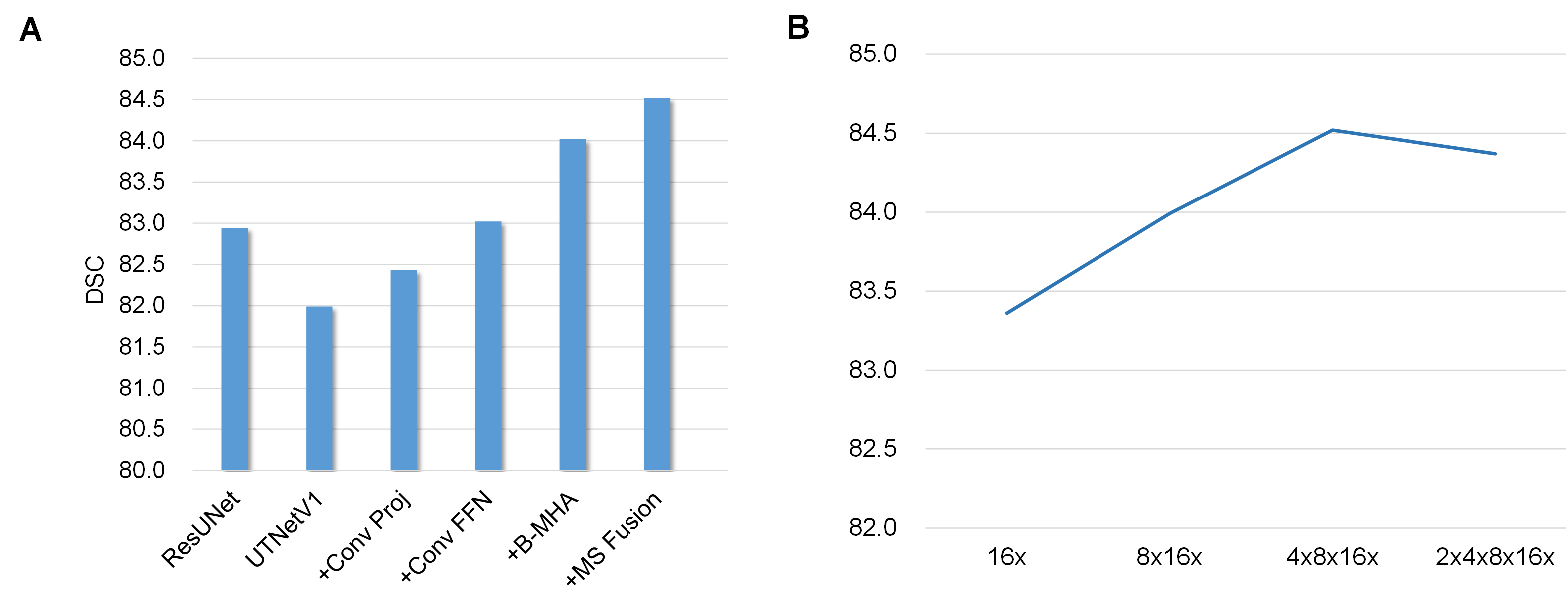}
    \caption{(A): Contribution from each component of MedFormer. (B): Ablation study on the position of B-MHA blocks. (C) Illustration of the initial semantic map generation.}
    \label{fig:sup_ablation}
\end{figure*}

Fig. \ref{fig:sup_ablation} presents a comprehensive ablation analysis of MedFormer by incrementally adding individual model components. Experiments are conducted on the BCV dataset in a 3D setting. First, we replace the middle and deep layers in ResUNet with efficient attention modules introduced in UTNet; this results in slight performance degradation due to the simple linear low-rank projection being unable to handle complex 3D anatomical structures. The performance increases after introducing convolutional projection and FFN, demonstrating the effectiveness of the convolutional inductive bias.

Next, we implement the proposed B-MHA, significantly improving the performance to 84.02, which indicates that the non-linear low-rank projection and continuous refinement can encode better global semantics. Finally, the multi-scale fusion module further boosts the performance to 84.52. We also attempted to include relative position embedding, but it did not yield a noticeable improvement. This may be due to the compact size of the B-MHA semantic map, rendering the relative position information less meaningful. The convolutional projection and FFN can introduce relative position to the B-MHA module.

In Fig. \ref{fig:sup_ablation} (B), we investigate the role of B-MHA blocks' position by replacing the residual blocks of ResUNet with the proposed B-MHA blocks at different levels. We find that applying B-MHA to the $16\times$ down-sampled feature map yields limited improvement compared to ResUNet. Applying B-MHA to a higher-resolution level gradually increases the performance up to $4\times$, while applying it to an even higher-resolution level decreases the performance. We hypothesize that $2\times$ features represent low-level textures, making the collection of long-range associations on such detailed features ineffective.

\subsection{Supplementary results on the collected large cardiac dataset.}

\begin{figure*}[]
    \centering
    \includegraphics[width=0.75\textwidth]{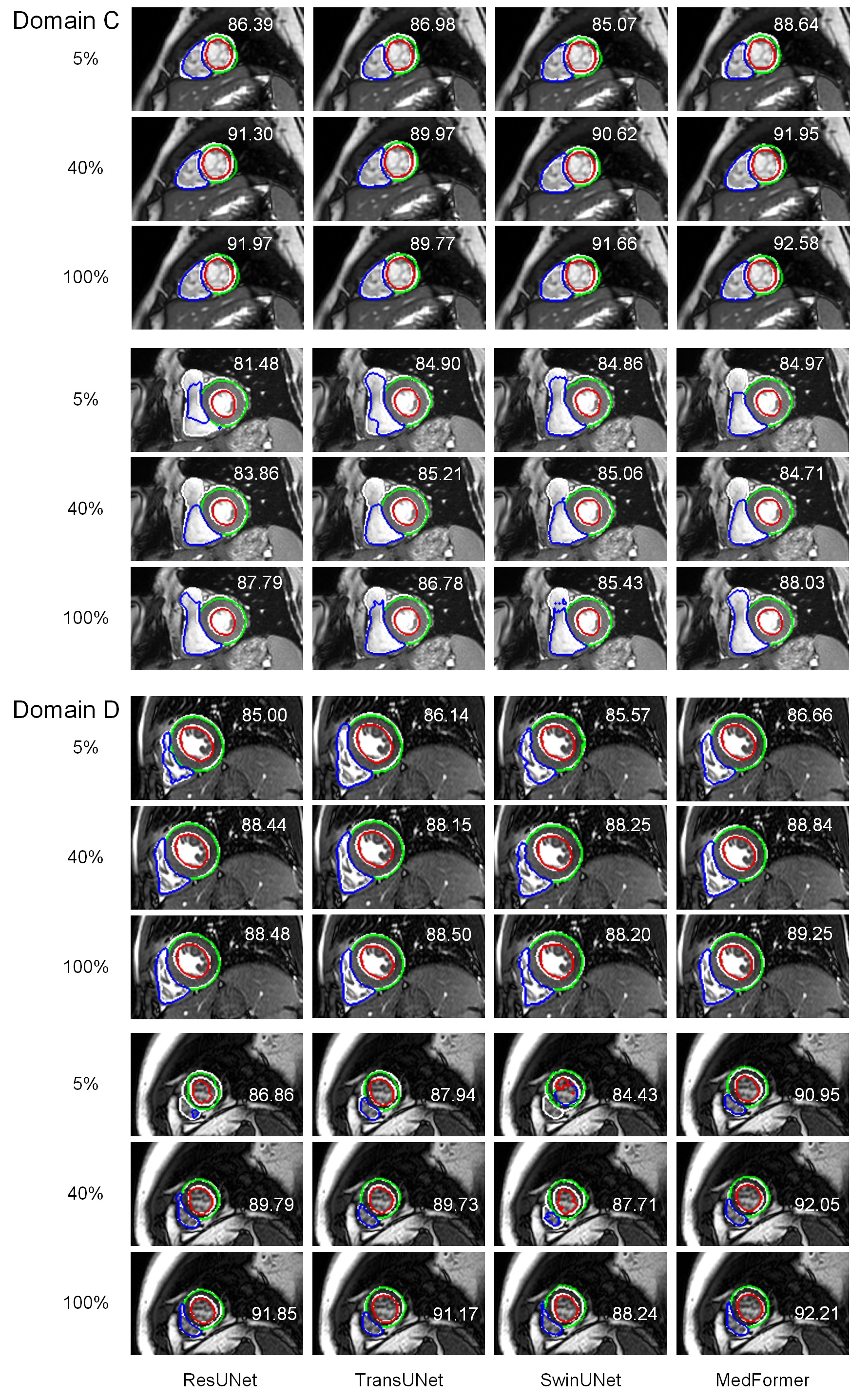}
    \caption{Generalization performance on unseen domains C and D from the collected cardiac dataset with 5\%, 40\% and 100\% training data ratios respectively. White: ground truth, Red: LV. Green: MYO. Blue: RV. The numbers on each image is the obtained Dice scores of the associated case. }
    \label{fig:sup_cardiac}
\end{figure*}

Table \ref{tab:sup_cardiac} shows the 95HD of all models across training ratios. MedFormer demonstrates superior performance on 95HD, particularly with 5 percent data input. The other results are consistent with the DSC analysis of the large cardiac dataset. Fig. \ref{fig:sup_cardiac} shows the segmentation boundary of models on the unseen domains C and D at 5\%, 40\%, and 100\% training data ratios. Overall, MedFormer has the best robustness against domain shifts.

\begin{table*}[]
    \centering
    \caption{95 HD on large cardiac MRI dataset with different training data ratios}
    \setlength{\tabcolsep}{5mm}{
    \begin{threeparttable}
    \begin{tabular}{c|c|c|c|c|c|c}
    \toprule
                Arch.   & Models &5\% & 10\% & 40\% & 70\% & 100\% \\ \hline \midrule
    \multirow{6}*{CNN}  & UNet \cite{ronneberger2015u}  & 7.34 & 6.59 & 5.62 & 5.61 & 5.22  \\ \cline{2-7}
                        & Attn UNet \cite{schlemper2019attention} & 7.69 & 5.87 & 5.92 & 5.29 & 5.22 \\ \cline{2-7}
                        & UNet++ \cite{zhou2018unet++} & 8.33 & 6.95 & 5.18 & 5.09 & 4.87 \\ \cline{2-7}
                        & ResUNet  & 7.37 & \textbf{5.54} & 4.97 & 4.86 & 5.1 \\ \cline{2-7}
                        & R50-UNet \cite{he2016deep}  & 8.01 & 6.01 & \textbf{4.56} & 4.44 & 4.49 \\ \cline{1-7}
    \multirow{6}*{TFM}  & TransUNet \cite{chen2021transunet}  & 6.85 & 5.66 & 5.42 &4.64 & 4.89\\ \cline{2-7}
                        & TransUNet$\dagger$  & 7.67 & 5.86 & 4.96  & 4.95 & 5.40 \\ \cline{2-7}
                        & SwinUNet \cite{cao2021swin}  & 16.06 & 9.2  & 6.71   & 6.47 & 6.27  \\ \cline{2-7}
                        & SwinUNet$\dagger$ & 6.09 & 5.81 & 5.64 & 5.34 & 5.72 \\ \cline{2-7}
                        & UTNet \cite{gao2021utnet}  & 7.05 & 5.63 & 4.99 &  4.82  & 4.92 \\ \cline{2-7}
                        & MedFormer & \textbf{5.82}  & 5.56  & 4.79 &  \textbf{4.43}  & \textbf{4.47}\\

    \bottomrule
   
    \end{tabular}
    
    \begin{tablenotes}
    \footnotesize
    \item $\dagger$ indicates the model is initialized with pre-trained weights on ImageNet.
    \end{tablenotes}
    
    \end{threeparttable}
    \label{tab:sup_cardiac}}
\end{table*}

\subsection{Supplementary results on seven public datasets}

Table \ref{tab:acdc_hd} presents the 95HD results on the ACDC dataset. MedFormer achieves top performance among all models in both 2D and 3D settings. 

\begin{figure*}[h]
    \centering
    \includegraphics[width=\textwidth]{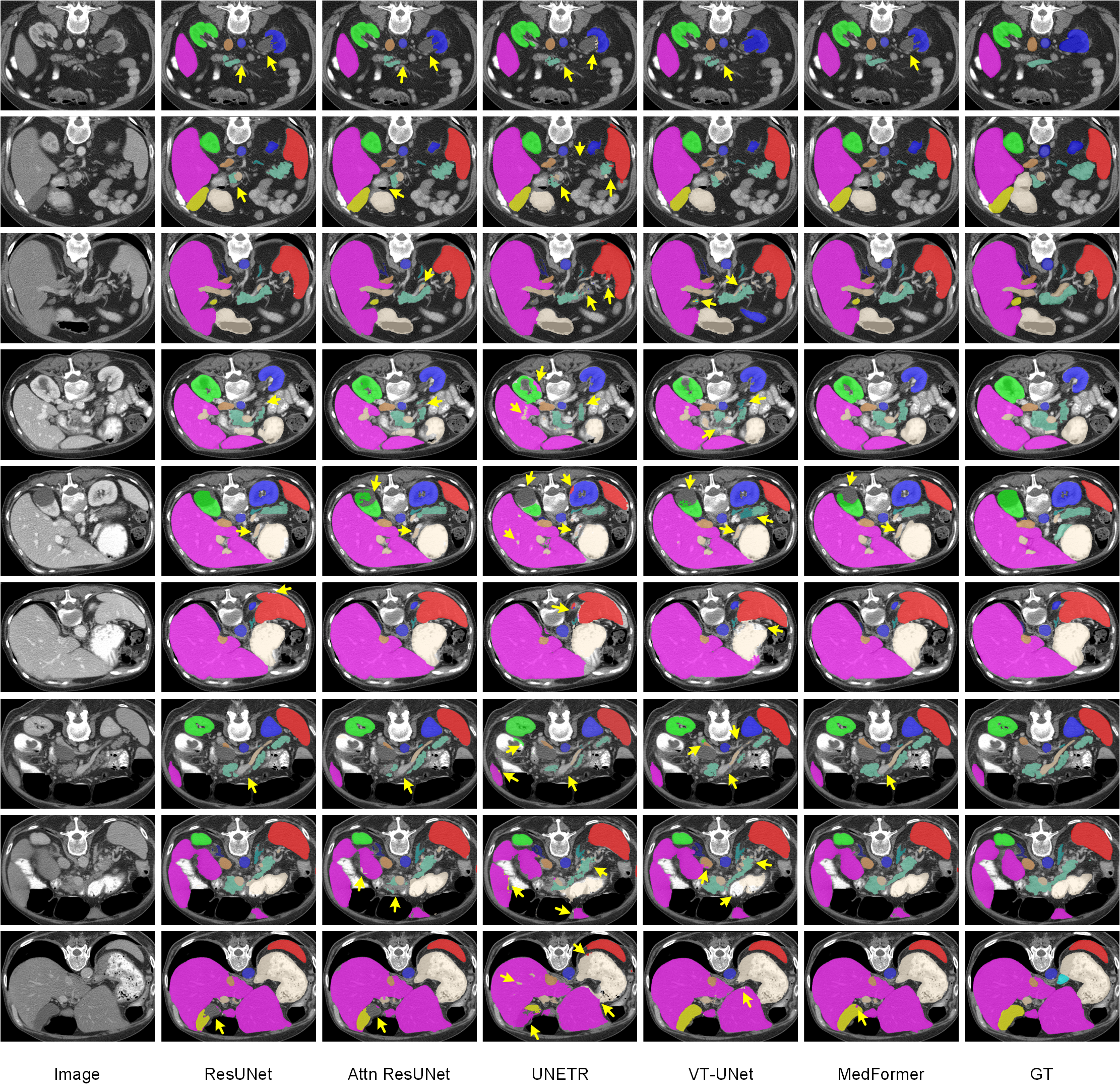}
    \caption{More visualizations on BCV dataset. Yellow arrows indicate the notable erroneous areas.}
    \label{fig:sup_bcv}
\end{figure*}

Table \ref{tab:sup_bcv_dsc} and Table \ref{tab:sup_bcv_hd} show the detailed DSC and 95HD on each organ in the BCV dataset. On the majority of organs, MedFormer has the highest performance in terms of both metrics. MedFormer not only achieves better results on organs with large volumes and clear boundaries (such as spleen, kidney, and liver), but also surpasses other methods on organs with complex shapes and small volumes (such as pancreas and veins). Fig. \ref{fig:sup_bcv} shows more segmentation visualizations. MedFormer has the most accurate delineation boundary. UNETR and VT-UNet suffer from limited data and are unable to precisely model the complex abdominal organs, tending to have mis-segmented areas and outliers. Note that our reported results of UNETR is lower than that of their original paper, since we report the results with five-fold cross-validation instead of the official test set. Moreover, we do not use any ensemble or additional data to gain potential optimistic results.

Table \ref{sup_tab:lits_kits} shows the Dice score of LiTS and KiTS. As the liver and kidney are large and have clear boundary, most models demonstrate good performance. The key differences lie in the tumor. MedFormer achieves a significant lead on both types of tumors. The LiTS dataset contains tumors of various shapes and sizes, with a dispersed distribution, making it more challenging. MedFormer's advantages become even more apparent in this context.

\begin{figure*}[h]
    \centering
    \includegraphics[width=\textwidth]{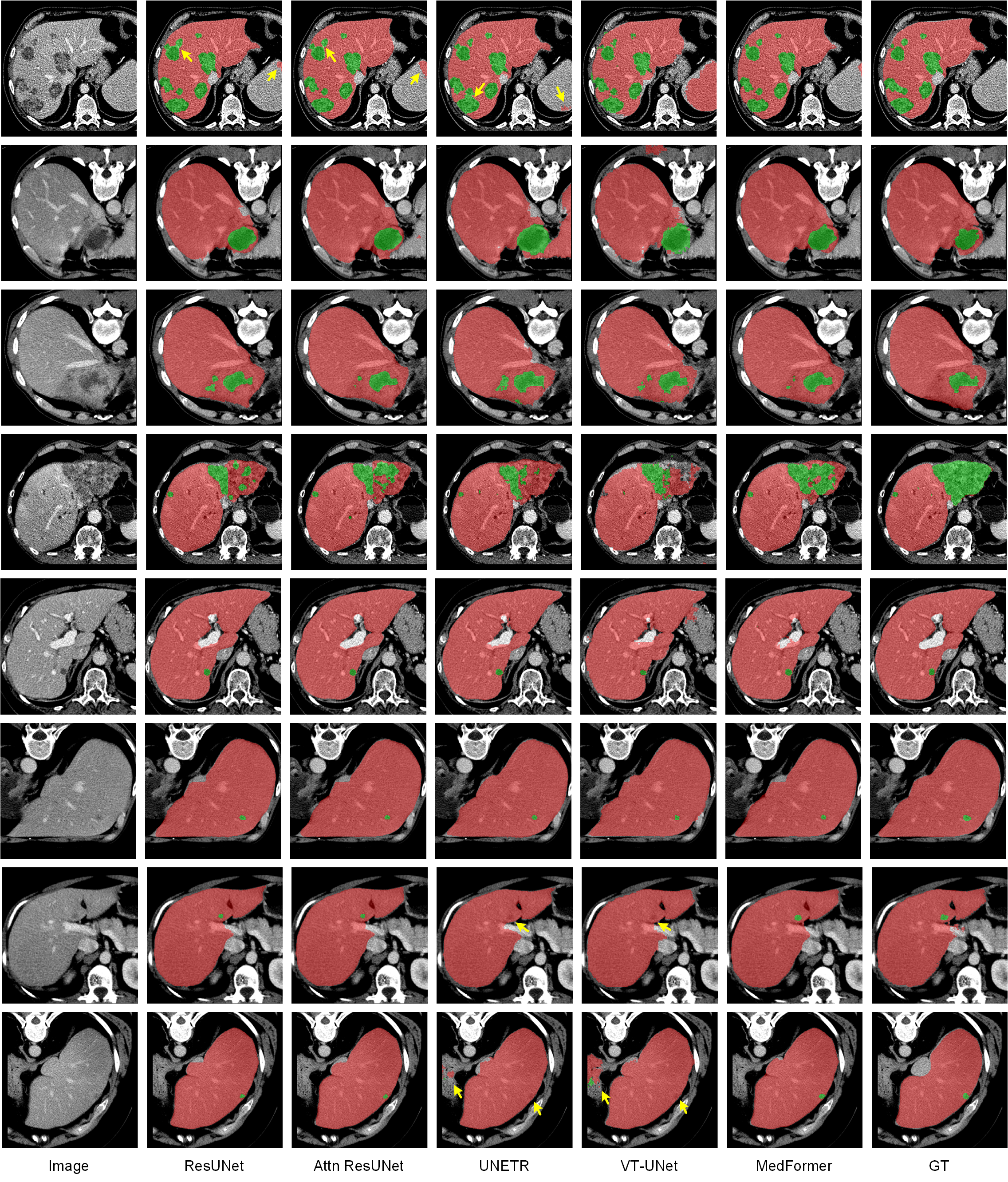}
    \caption{More visualizations on LiTS dataset. Yellow arrows indicate the notable erroneous areas.}
    \label{fig:sup_lits}
\end{figure*}

Fig. \ref{fig:sup_lits} presents additional visualizations on the LiTS dataset. MedFormer consistently demonstrates the highest accuracy for both large and small tumors. The first four rows depict cases with large tumors, where most models produce satisfactory segmentation results. Rows 6-7 showcases with small tumors and high contrast; MedFormer outperforms all other models in these cases. The last two rows display small tumors with low contrast, in which both UNETR and VT-UNet fail to detect the tumors. These examples highlight the limitations of ViT-like structures for detailed segmentation. Specifically, UNETR first divides images into $16\times 16\times 16$ patches, flattens them, and linearly embeds them into tokens. This process is equivalent to a $16\times$ down-sampling. The tokens are then processed using self-attention for context modeling. However, the $16\times 16\times 16$ patches are too coarse for fine-grained structures. For instance, tiny tumors smaller than $16\times 16\times 16$ are flattened along with the surrounding background into a single token, discarding all structural and positional information. We found that modeling such tiny, low-contrast tumors using MLPs in the self-attention module is highly ineffective. In contrast, the proposed efficient B-MHA allows MedFormer to encode visual representations in $4\times$, $8\times$, and $16\times$ token maps hierarchically. The proposed multi-scale fusion module further refines representations across scales. Consequently, MedFormer exhibits sensitivity to fine-grained details and successfully detects tiny tumors.

\begin{table*}[h]
    
    \centering
    \caption{DSC results on the ACDC dataset.}
        \setlength{\tabcolsep}{4mm}{
    \begin{threeparttable}
    \begin{tabular}{c|c|c|c|c}
        \toprule
         Model & RV & MYO & LV & AVG \\ \midrule
               
        UNet   & $90.10\pm 1.21$ &$88.86\pm 0.32$  & $94.17\pm 1.07$  & $91.05\pm 0.69$ \\\hline
        UNet++ & $90.00\pm 1.19$  & $89.31\pm 0.43$  &$94.33\pm1.01$ &$91.21\pm 0.72$ \\ \hline
        ResUNet & $90.44\pm 1.08$ & $89.13\pm 0.43$  & $94.16\pm 1.09$  & $91.25\pm 0.66$ \\ \hline
        Attn UNet & $90.48\pm 1.08$  & $88.93\pm 0.33$ & $94.13\pm 1.36$  &$91.18\pm0.80$  \\ \hline
        TransUNet & $89.66\pm 1.46$  & $88.07\pm 0.39$  & $93.98\pm 0.99$ & $90.57\pm 0.74$   \\\hline
        TransUNet $\dagger$& $89.78\pm 1.29$  & $88.74\pm 0.58$ & $94.03\pm 1.29$ & $90.85\pm 0.97$   \\\hline
        SwinUNet & $81.81\pm 3.27$& $81.18\pm 2.86$& $90.01\pm 2.50 $  & $84.34\pm 2.77$ \\\hline
        SwinUNet $\dagger$ & $90.34\pm 1.11$ & $88.70\pm 0.47$  & $94.05\pm 1.40$  & $91.03\pm 0.91$  \\\hline
        UTNet & $90.41\pm 1.03$  & $89.15\pm 0.58$  & $94.39\pm 1.24$ & $91.32\pm 0.81$ \\\hline
        MedFormer & \bm{$90.77\pm 0.95$}  & \bm{$89.84\pm 0.45$}   & \bm{$94.71\pm 1.07$}  & \bm{$91.77\pm 0.71$}  \\ \hline \midrule
        
        ResUNet & $89.77\pm 0.29$  & $88.95\pm 0.17$  & $95.16\pm 0.22$  & $91.30\pm 0.21$ \\ \hline
        Attn ResUNet & $90.01\pm 0.20$  & $89.09\pm 0.16$ & $95.22\pm 0.09$  &$91.44\pm 0.08$  \\ \hline
        UNETR & $84.52\pm 0.26$  & $84.36\pm 0.37$ & $92.57\pm 0.17 $ & $87.15\pm 0.19$ \\ \hline
        VT-UNet  & $80.44\pm 2.92$ & $80.71\pm 2.07$ & $89.53\pm 1.99$  & $83.56\pm 2.32$  \\ \hline
        VT-UNet $\dagger$  & $89.44\pm 0.19$  & $88.42\pm 0.07$  & $95.53\pm 0.03$  & $91.13\pm 0.08$   \\ \hline
        nnFormer & $89.14 \pm 1.21$ & $87.01 \pm 0.58$ & $94.22 \pm 0.76$ & $90.12 \pm 0.67$ \\ \hline
        MedFormer & \bm{$90.95\pm 0.28$}  & \bm{$89.71\pm 0.08$}  & \bm{$95.76\pm 0.06$}  & \bm{$92.14\pm 0.12$} \\ \bottomrule

    \end{tabular}
    
    \begin{tablenotes}
    \footnotesize
    \item The top table shows the results of 2D models, and the bottom table shows 3D models.\\
     \item $\dagger$ indicates the model is initialized with pre-training weights on ImageNet21K.
    \end{tablenotes}
    
    \end{threeparttable}
    
    \label{tab:acdc}}
\end{table*}

\begin{table*}[h]
    
    \centering
    \caption{95 HD on the ACDC dataset.}
    \setlength{\tabcolsep}{5mm}{
    \begin{threeparttable}
    \begin{tabular}{c|c|c|c|c}
        \toprule
         Model & RV & MYO & LV & AVG\\\midrule
        UNet   &$5.91\pm 1.37$  & $2.49\pm 0.50$  & $2.94 \pm 0.76$ & $3.78\pm 0.74$ \\\hline
        UNet++  & $6.91\pm 2.31$  & \bm{$2.42\pm 0.31$}  & \bm{$2.83\pm 0.45$}  & $4.06\pm 0.79$\\ \hline
        ResUNet & $5.53\pm 1.00$ & $2.53\pm 0.39$  &$3.08\pm 0.85$ & $3.71\pm 0.62$\\ \hline
        Attn UNet  & $5.48\pm 0.92$  &$2.82\pm 0.77$  & $3.25\pm 1.31$  & $3.85\pm 0.91$ \\ \hline
        TransUNet  & $6.22\pm 1.46$  & $3.08\pm 1.21$  & $3.28\pm 1.42$ & $4.19\pm 1.08$  \\\hline
        TransUNet $\dagger$  & $6.31\pm 1.32$  & $3.06\pm 1.31$  & $3.38\pm 1.45$  & $4.25\pm 1.29$   \\\hline
        SwinUNet & $13.51\pm 2.35$  & $6.54\pm 3.44$ & $7.80\pm 3.60$  & $ 9.28\pm 2.82$\\\hline
        SwinUNet $\dagger$ & $5.44\pm 1.05$ & $3.11\pm 1.39$  & $3.78\pm 2.21$  & $4.11\pm 1.52$ \\\hline
        UTNet  & $5.59\pm 0.95$  &$2.54\pm 0.57$ & $2.96\pm 1.12$  & $3.69\pm 0.79$\\\hline
        MedFormer  & \bm{$5.27\pm 0.54$}  & $2.49\pm 0.48$   & $2.91\pm 0.85$  & \bm{$3.55\pm 0.54$} \\ \hline \midrule
        
        ResUNet  & $5.04\pm 0.27$ & $2.54\pm 0.26$ &$2.77\pm 0.20$  & $3.46\pm 0.23$\\ \hline
        Attn ResUNet & $5.05\pm 0.32$  & $2.39\pm 0.18$  &$2.73\pm 0.08$ & $3.39\pm 0.15$ \\ \hline
        UNETR &  $9.06\pm 3.08$  & $3.30\pm 1.89$ & $4.10\pm 0.45$  & $5.48\pm 1.13$\\ \hline
        VT-UNet & $11.09\pm 1.78$ & $5.24\pm 2.02$ & $6.32\pm 2.95$ & $5.55\pm 2.25$  \\ \hline
        VT-UNet $\dagger$   & $5.02\pm 0.18$ & \bm{$2.14\pm 0.02$}  & \bm{$2.19\pm 0.05$}  & $3.12\pm 0.07$  \\ \hline
        nnFormer & $5.61 \pm 0.79$ & $3.02 \pm 0.27$ & $3.14 \pm 0.36$ & $3.93 \pm 0.38$ \\ \hline
        MedFormer  & \bm{$4.47\pm 0.32$}  & $2.23\pm 0.10$  & $2.55\pm 0.07$  & \bm{$3.08\pm 0.11$}\\ \bottomrule
               
    \end{tabular}
        
        \begin{tablenotes}
    \footnotesize
    \item The top table shows the results of 2D models, and the bottom table shows 3D models.\\
     \item $\dagger$ indicates the model is initialized with pre-training weights on ImageNet21K.
    \end{tablenotes}
    \end{threeparttable}
    \label{tab:acdc_hd}}

\end{table*}

\begin{table*}[h]
    \centering
    \caption{DSC of 13 organs in the BCV dataset.}
    \setlength{\tabcolsep}{2mm}{
    \begin{threeparttable}
    \begin{tabular}{c|c|c|c|c|c|c|c|c}
    \toprule
    DSC & nnUNet$\ast$ &ResUNet & At-UNet & UNETR & SwinUNETR & VT-UNet$\dagger$ & nnFormer & MedFormer \\
    \midrule
    Spleen &90.83 & 92.41 & 92.58 & 90.82 & 92.49 & 91.12 & 90.73 &\textbf{93.15} \\
    R Kid  & 89.39 &91.22 & \textbf{91.52} & 87.11 & 88.26 &87.14 & 87.1 & 91.21 \\
    L Kid & 86.75 & 91.79 & 91.69 & 83.01 & 86.87 & 84.43 & 73.63 &\textbf{92.40}\\
    Gall & 66.32 & 67.14 & 71.51 & 65.56 & 66.53 & 70.07 & 73.63 &\textbf{77.36}\\
    Eso & \textbf{78.40} & 77.31 & 77.75 & 70.52 &76.18 & 73.69& 76.33 &78.05\\
    Liv & 95.57 & 95.50 & 95.57 & 94.42 & 95.65 & 94.79& 94.35 & \textbf{95.85} \\
    Sto & 88.16 & 88.37 & 87.29 & 81.11 & 82.87 & 82.16& 84.28 &\textbf{90.23} \\
    Aor & 92.29 & 91.64 & 91.63 & 86.79 & 91.04 & 88.54& 89.24 & \textbf{92.43} \\
    IVC & 86.38 & 85.63 & 86.05 & 80.80 & 84.73 & 81.07& 85.24 & \textbf{86.47} \\
    Veins & \textbf{76.59} & 74.33 &74.11 & 66.05 & 71.70 & 68.49 & 71.28 & 76.17\\
    Pan & 81.76 & 80.65 & 79.49 & 68.08 & 73.99 & 71.47& 75.61 & \textbf{81.92}\\
    R AG & 71.48 & 71.13 & 71.50 & 66.83 & \textbf{71.79} & 68.43 & 68.55 & 71.18\\
    L AG & 72.38 & 71.12 & 70.06 & 63.84 & 68.17 & 65.85 & 67.85 & \textbf{72.39} \\ \midrule
    AVG & 82.79 & 82.94 & 83.13 & 77.30 & 80.79 & 79.02 & 80.87 & \textbf{84.52}\\
    \bottomrule
    \end{tabular}
    \begin{tablenotes}
    \footnotesize
    \item $\ast$ means the result is obtained from models trained using nnUNet framework (3D\_fullres). Others are trained with our framework.
     \item $\dagger$ indicates the model is initialized with pre-training weights on ImageNet21K.
    \end{tablenotes}
    \end{threeparttable}
    \label{tab:sup_bcv_dsc}}
\end{table*}

\begin{table*}[h]
    \centering
    \caption{95HD of 13 organs in the BCV dataset.}
    \setlength{\tabcolsep}{2mm}{
    \begin{threeparttable}
    \begin{tabular}{c|c|c|c|c|c|c|c|c}
    \toprule
    95HD & nnUNet$\ast$& ResUNet & At-UNet & UNETR & SwinUNETR & VT-UNet$\dagger$ & nnFormer & MedFormer \\
    \midrule
    Spleen &-& 11.54 & 11.048 & 29.42 & 29.22 & 30.28& 12.65 &\textbf{4.95} \\
    R Kid  &-& \textbf{9.05} & 20.08 & 19.65 & 12.10 &18.39 & 13.68 & 10.05 \\
    L Kid &-& 4.52 & 5.77 & 28.41 & 15.84 & 28.58& 23.84 &\textbf{3.64}\\
    Gall &-& 65.95 & 34.63 & 33.45 & 31.52 & 42.29 & 21.97 &\textbf{25.98}\\
    Eso &-& 8.37 & 11.21 & 21.29 &12.70 & 9.80& 8.29 &\textbf{7.78}\\
    Liv &-& 24.9 & 22.60 & 29.61 & 20.30 & 21.85& 15.92 & \textbf{16.38} \\
    Sto &-& 30.05 & 32.24 & 46.4 &34.22 & 35.88& 27.88 &\textbf{29.4} \\
    Aor &-& 7.32 & 7.82 & 36.13 &12.89 & 18.9& 11.38 &\textbf{6.47} \\
    IVC &-& \textbf{9.93} & 13.28 & 39.66 &13.05 &14.27 & 10.59 & 11.96 \\
    Veins &-& 15.32 &12.21 & 20.22 &14.51& 17.41& 15.09 &\textbf{12.05}\\
    Pan &-& \textbf{8.38} & 9.15 & 25.52 & 21.31 &20.04 & 16.69 & 12.23\\
    R AG &-& 4.60 & \textbf{4.39} & 6.57 & 5.68 & 7.01& 6.44 & 9.57\\
    L AG &-& 5.65 & 8.64 &  15.21 &11.08 & 12.77 & 9.02 & \textbf{5.43} \\ \midrule
    AVG &-& 15.89 & 14.85 & 27.04 & 18.03 & 21.34 & 14.88 &\textbf{11.99}\\
    \bottomrule
    \end{tabular}
    \begin{tablenotes}
    \footnotesize
    \item $\ast$ means the result is obtained from models trained using nnUNet framework (3D\_fullres). Others are trained with our framework.
     \item $\dagger$ indicates the model is initialized with pre-training weights on ImageNet21K.
    \end{tablenotes}
    \end{threeparttable}
    \label{tab:sup_bcv_hd}}
\end{table*}

\begin{table*}[h]
    
    \centering
    \caption{DSC results on the LiTS and KiTS dataset.}
        \setlength{\tabcolsep}{4mm}{
    \begin{threeparttable}
    \begin{tabular}{c|ccc|ccc}
        \toprule
         Model & Liver & Tumor & LiTS AVG & Kidney & Tumor & KiTS AVG \\ \midrule
               
        nnUNet & 95.76 & 62.53 & 79.14 & \bf97.02 & 83.67 & 90.34\\\hline
        ResUNet & 95.70 & 63.42	&79.56&	96.75	&83.28&	90.01 \\\hline
        Attn ResUNet  &95.52&	64.07&	79.79&	96.74&	83.63&	90.18 \\ \hline
        UNETR &93.23&	53.02&	73.12&	94.92&	70.52&	82.72\\\hline
        SwinUNETR&95.01	&61.1	&78.05&	96.11&	78.55&	87.33\\\hline
        VT-UNet $\dagger$   &94.65&	53.14&73.89	&95.6&	78.82&	87.21\\ \hline
        nnFormer & \bf95.81&	62.95&	79.38	&96.15&	80.69&	88.42\\ \hline
        MedFormer & 95.66&	\bf68.06&	\bf81.86	&96.86&\bf84.47&	\bf90.67\\ \bottomrule

    \end{tabular}
    
    \begin{tablenotes}
    \footnotesize
    \item The top table shows the results of 2D models, and the bottom table shows 3D models.\\
     \item $\dagger$ indicates the model is initialized with pre-training weights on ImageNet21K.
    \end{tablenotes}
    
    \end{threeparttable}
    
    \label{sup_tab:lits_kits}}
\end{table*}

\begin{table*}[h]
    \centering
    \caption{DSC of 15 organs in the AMOS CT dataset.}
    \setlength{\tabcolsep}{2mm}{
    \begin{threeparttable}
    \begin{tabular}{c|c|c|c|c|c|c|c|c}
    \toprule
DSC	&nnUNet*	&ResUNet&	Attn UNet&	UNETR	&SwinUNETR	&VT-UNet$\dagger$	&nnFormer&	MedFormer \\\hline
SPL	&96.31	&96.29	&96.32	&92.68	&95.49	&95.3	&95.91	&\bf 97.13\\\hline
R Kid	&95.29	&95.95	&95.97	&88.46	&93.82	&94.27	&93.51	&\bf96.69\\\hline
L Kid	&96.28	&95.95	&95.88	&90.57	&94.47	&96.43	&94.80	&\bf96.70\\\hline
GBL	&81.53	&87.01	&87.73	&66.5	&77.34	&78.2	&78.47	&\bf88.53\\\hline
ESO	&85.72	&84.16	&84.52	&73.31	&83.05	&75.49	&81.09	&\bf85.73\\\hline
Liv	&97.05	&97.08	&96.95	&94.11	&95.95	&96.42	&95.89	&\bf97.78\\\hline
STO	&90.77	&92.45	&92.59	&78.73	&88.94	&87.49	&89.40	&\bf93.38\\\hline
AOR	&95.37	&94.59	&94.6	&91.37	&94.66	&93.85	&94.16	&\bf95.39\\\hline
IVC	&\bf91.53	&90.68	&90.57	&83.99	&89.58	&85.22	&88.25	&91.43\\\hline
PAN	&87.39	&86.85	&86.91	&74.49	&84.91	&81.68	&85.00	&\bf88.05\\\hline
RAG	&79.83	&79.01	&79.15	&68.15	&77.2	&72.22	&75.04	&\bf80.19\\\hline
LAG	&\bf81.12	&79.65	&80.08	&65.28	&78.35	&70.89	&75.92	&80.51\\\hline
Duo	&82.56	&83.32	&84.01	&62.35	&78.59	&74.61	&78.45	&\bf85.17\\\hline
BLA	&88.42	&88.93	&88.56	&77.44	&85.79	&81.36	&83.91	&\bf90.38\\\hline
Pro	&83.81	&81.92	&82.54	&67.52	&77.39	&72.48	&74.58	&\bf84.55\\\hline
Avg	&88.87	&88.92	&89.09	&78.33	&86.37	&83.73	&85.63	&\bf90.11\\\hline
    \bottomrule
    \end{tabular}
    \begin{tablenotes}
    \footnotesize
    \item $\ast$ means the result is obtained from models trained using nnUNet framework (3D\_fullres). Others are trained with our framework.
     \item $\dagger$ indicates the model is initialized with pre-training weights on ImageNet21K.
    \end{tablenotes}
    \end{threeparttable}
    \label{tab:sup_bcv_dsc}}
\end{table*}

\bibliographystyle{IEEEtran}
\bibliography{reference.bib}
%


